\documentclass[preprint]{revtex4-1}

\usepackage{graphicx}

\usepackage{amssymb,amsfonts,amsmath}
\usepackage{subfig}

\begin{document}

%%%%%%%%%%%%%%%%%%%%%%%%%%%%%%
%\title{Modeling collective human mobility in urban areas}
%\title{Understanding the exponential law of intra-urban mobility}
\title{Modeling collective human mobility: Understanding exponential law of intra-urban movement}

\author{Xiao Liang}
\email{liangxiao@nlsde.buaa.edu.cn}
\author{Jichang Zhao}
\email{zhaojichang@nlsde.buaa.edu.cn}
\author{Li Dong}
\email{donglixp@gmail.com}
\author{Ke Xu}
%\thanks{Corresponding author}
\email[Corresponding author: ]{kexu@nlsde.buaa.edu.cn}
\affiliation{State Key Lab of Software Development Environment, Beihang University}

%%%%%%%%%%%%%%%%%%%%%%%%%%%%%%%%%%%%%%%%%%%%%%%%%%%%%%%%%%%%%%%%

\begin{abstract} 
It is very important to understand urban mobility patterns because
most trips are concentrated in urban areas. In the paper, a new model
is proposed to model collective human mobility in urban areas. The
model can be applied to predict individual flows not only in intra-city but
also in countries or a larger range. Based on the model, it can be
concluded that the exponential law of distance distribution is
attributed to decreasing exponentially of average density of human
travel demands. Since the distribution of human travel demands only
depends on urban planning, population distribution, regional functions
and so on, it illustrates that these inherent properties
of cities are impetus to drive collective human movements.
\end{abstract}


\maketitle

Understanding human movement patterns is considered as  a long-term
challenging work for a long time. It is very crucial to urban planning
\cite{Rozenfeld2008b, Jing2012, Zheng2011}, 
epidemics spreading \cite{Balcan22122009,Colizza2007,Wang2012} and traffic
engineering \cite{Viboud21042006,Jung2012,Goh2012}.
During the past few years, various mobile devices  (e.g. cellphones and GPS navigators)
that support geolocation have been widely used in our daily life. As
proxies, these devices record massive amounts of individual
tracks. Benefited from it, the research of human mobility has 
attracted more and more attention of scientists from multiple
disciplines such as physics, computer science and
biology.

In recent studies, Brockmann et al. \cite{Brockmann2006} discovered
human travel displacements can be described by a power-law
distribution by investigating the dispersal of bank notes in the
United States. Gonz\'{a}lez et al. \cite{Gonzalez2008} studied mobility patterns of mobile phone
users in European countries and found that their travel distances are
distributed according to a truncated power-law. Moreover,  the
similar scaling law was also observed in \cite{Song2010a} and
\cite{Jiang2009} separately. Therefore, in order to understand the cause of  the
scaling law, some researchers tried to propose possible explanations
from individual movement viewpoint \cite{Han2011,Hu2011a}. It is worthy to note that these researches
characterized human travel occurred in large scale
of space, including trips from countries to countries or cities to cities.

Furthermore, there are also many studies which focus on human movement in urban
areas. For example, trajectories of passengers by taxis were investigated
separately in three cities: Lisbon \cite{Veloso2011}, Beijing \cite{Liang2012} and
Shanghai \cite{Peng2012}. And the three studies all suggested that trip distances obey
exponential distributions rather than power-law
ones. Bazzani et al. \cite{Bazzani2010} analyzed daily round-trip lengths of private
cars' drivers in Florence and revealed an exponential
law of lengths too. In addition, the distances of individuals'
movement in the London subway were found obviously deviating from the
power-law distribution as well \cite{Roth2011}. A more convincing
evidence is that exponential distributions of intra-urban travel
distances were demonstrated respectively in eight cities of 
Northeast China by analyzing the mobile phone data \cite{Kang2012},
which was not restricted to means of transportation. Even though a lot
of empirical studies, the understanding of intra-urban mobility is
still limited and there are no reasonable model to account for the
exponential law to the best of our knowledge.

In order to understand the exponential law of collective human
mobility in urban areas, it is essential to model 
individual flows from one region to the other in a city.
As we know, the gravity model \cite{Barthelemy2010} has already been
applied widely to predict flows, including human travel
\cite{Balcan22122009,Jung2012}, cargo ship movement
\cite{Kaluza06072010} and telephone communications
\cite{Krings2009}. Assuming $T_{ij}$ is the flux of individual between
location $i$ (with population $P_i$) and location $j$ (with population
$P_j$) and $d_{ij}$ is the distance between the two locations, a general
gravity law \cite{Simini2012a} can be represented by 
\begin{equation}T_{ij}=K\frac{P_i^\alpha P_j^\beta}{f(d_{ij})}\end{equation}\label{eqn:gravity}
where $K$, $\alpha$ and $\beta$ are tunable parameters and $f(d_{ij})$
is often selected as a power or exponential function of
$d_{ij}$. Especially, the gravity model with $\alpha=\beta=1$ can be derived from
entropy maximization \cite{Wilson-67}. Despite the prevailing gravity model, it still has some
disadvantages \cite{Simini2012a}. Particularly, the gravity model is
incompetent to explain the discrepancy of the numbers of individual flows in
both directions between a pair of locations. Consequently, Simini et
al. \cite{Simini2012a} put forward the radiation model without
parameters:
\begin{equation}\langle T_{ij}\rangle = T_i\frac{P_i P_j}{(P_i+P_{ij})(P_i+P_j+P_{ij})}\end{equation}\label{eqn:rad}
where $\langle T_{ij}\rangle$ is the expected flux from $i$ to $j$, 
$T_i$ is the number of trips started from $i$ and $P_{ij}$ is total
population of locations (except $i$ and $j$) from which to $i$ the distance
is less than or equal to $d_{ij}$. The model can predict population
movement between cities or countries successfully, but it is not
clear whether the model applies to intra-urban movement as well.

Therefore, by exploring human travels by taxis in Beijing, it is aimed
to figure out the answers to the following questions
in the paper:
\begin{itemize}
\item Whether can the radiation model predict intra-urban human flows?
  If not, how to model human mobility in urban areas.
\item What is the origin of the exponential law in intra-urban human
  mobility? Why do the distributions of travel distances in urban
  areas disagree with the ones at a larger scale? What is the inherent
  impetus to drive collective human movement?
\end{itemize}

\section{Empirical analysis of taxis' GPS data}
\subsection{Data description}
Here, we use the taxis' GPS data generated by over 10000 taxis in Beijing,
China, during three months ended on Dec. 31st, 2010 \cite{Liang2012}. From taxis'
locations and statuses of occupation (with passengers or without
passengers), trajectories of passengers can be observed. In order to
study intra-urban human mobility patterns, a total of 12009383
individuals' tracks were collected which occurred inside the 6th Ring
Road of Beijing.

\begin{figure}[htbp]
\centering
\includegraphics[scale=.5]{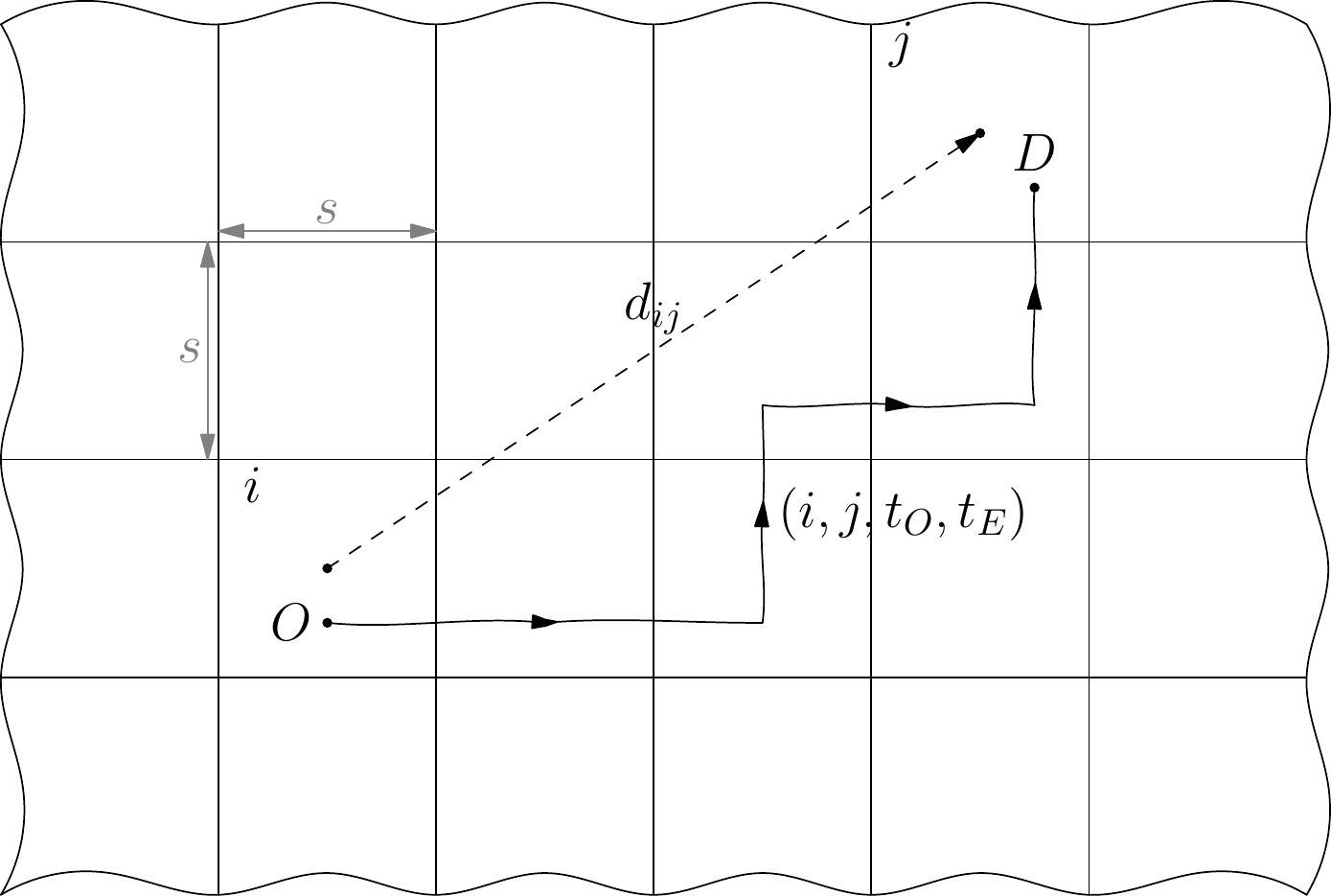}
\caption{Illustration of the grid division on a
  map. The cell size is $s\times s$. The solid line shows an
  individual trajectory from cell $i$ to $j$, which can be denoted by
  the tuple $(i, j, t_O, t_D)$. And the distance $d_{ij}$ between cell $i$
  and $j$ is described by the dashed line.}\label{fig:grid}
\end{figure}

For the purpose of investigating individual flows between regions in
urban areas, 
%Because there are a very large number of origins (pick-up points) and
%destinations (drop-off points) in our dataset
%and they seldom have exactly the same coordinates, 
the urban area in a map can be divided into discrete
grid-like cells of size $s\times s$ (selection of
$s$ will be described below). The number of
cells is $N$ and the Euclidean distance between the centers of cell $i$ and $j$
is defined as $d_{ij}$. Therefore, pick-up points (origins) and
drop-off points (destinations) of trajectories can be simplified to
cells which they lie in. 
As illustrated in Fig. \ref{fig:grid}, an individual trip
can be represented by a tuple $(l_O, l_D, t_O, 
t_D)$, where $l_O$ and $l_D$ are the origin and destination
cells ($l_O, l_D \in \{1, 2, \ldots, N\}$), $t_O$ and $t_D$
are the departure and arrival times. 

\subsection{Choice of cell size}
\begin{figure}[htbp]
\centering
\subfloat[] {
  \includegraphics[scale=.4]{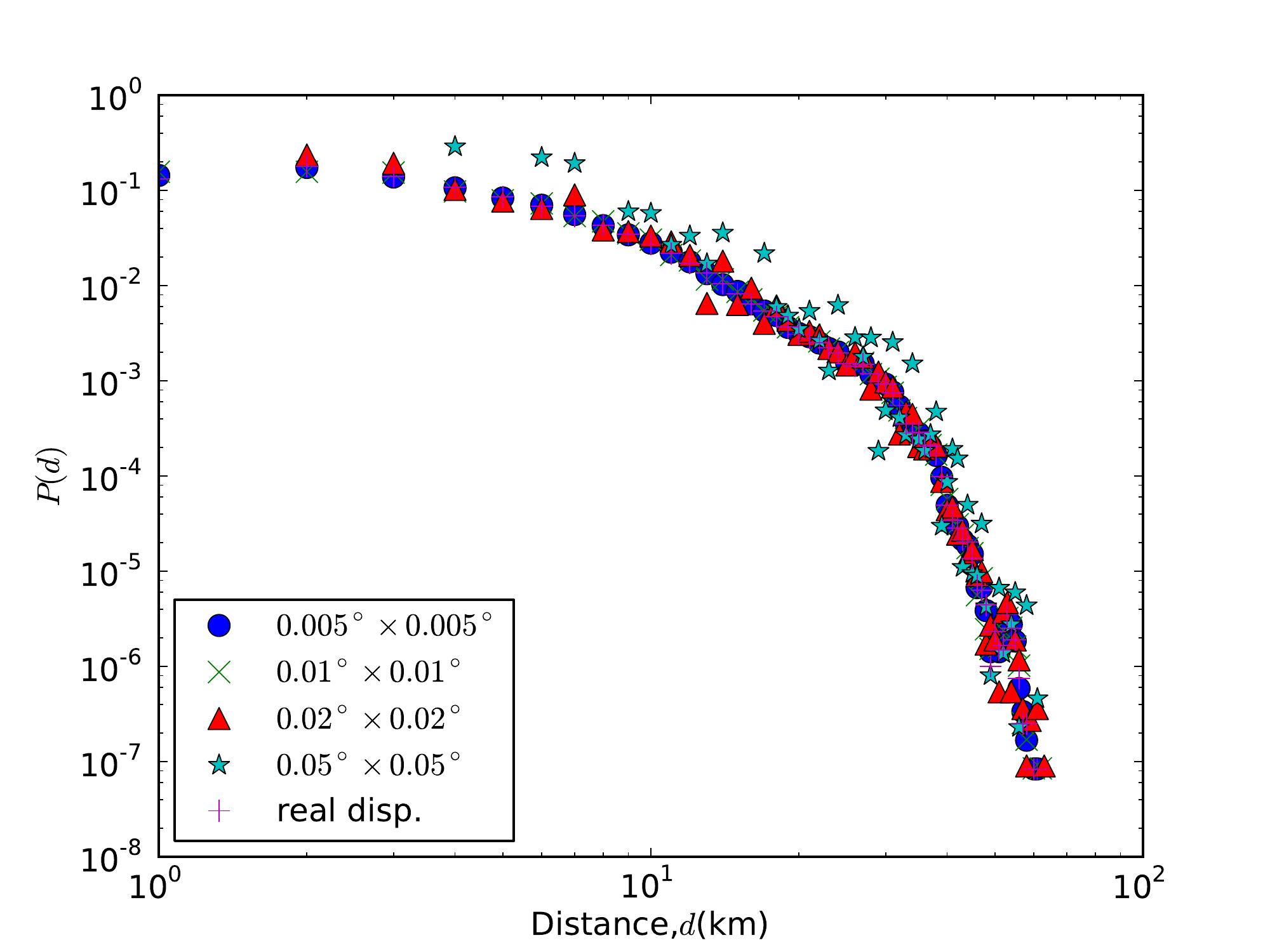}
  \label{fig:dif_size_disp}
}
\subfloat[] {
  \includegraphics[scale=.4]{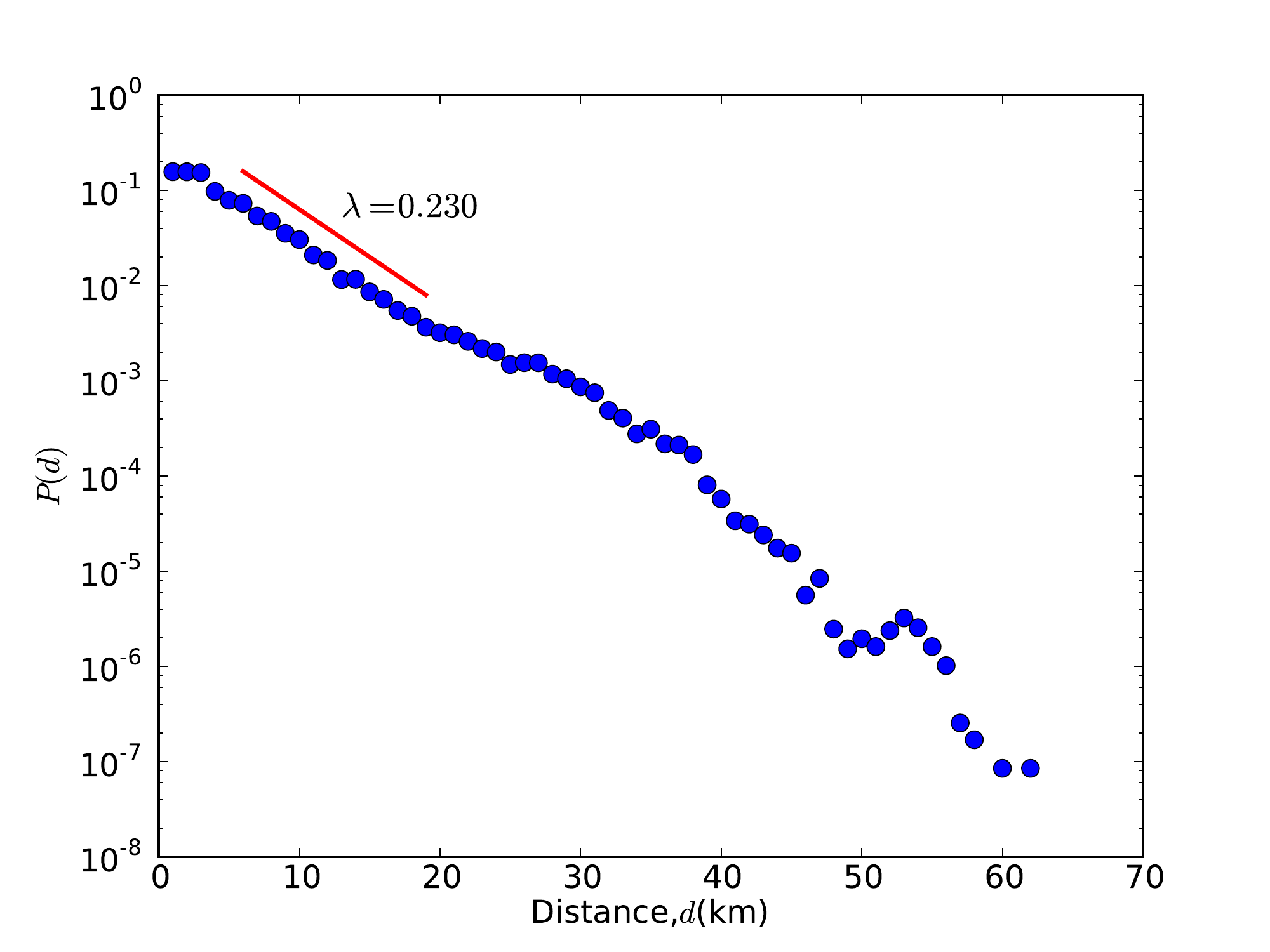}
  \label{fig:cell_disp}
}
\caption{Distributions of trip distances. (a) The comparison of
  distributions between approximated distances in different cell sizes
  and actual distances. (b) The distribution of approximated distances when cell size
  $s$ is $0.01^{\circ}$.}
\label{fig:dif_size}
\end{figure}

After dividing the urban areas into lattices, a trip length can be
approximated by the distance between centers of cells which the pick-up and
drop-off points lie in. In fact, it can be imagined that the location errors will
become more and more large with increasing the cell size. At the same
time, too small cell size is also undesirable because it would not reflect
the regular mobility patterns between different regions obviously and increase
computing costs. Therefore, it is important to choose an appropriate
cell size to model urban mobility better. As shown in
Fig. \ref{fig:dif_size_disp}, when cell size is larger than 
0.01 degree, there is some deviation from the real distribution of
trip displacements. So the cell size $s$ is determined as 0.01 degree in the
following paper.

For the cell size $s$ is $0.01^{\circ}$, the probability distribution
of approximated distances is
plotted in Fig. \ref{fig:cell_disp}. Bescause the fraction of trips
whose distances are less than 20km can reach nearly 98\%, the
curve can be fitted very well by an exponential function $y =
Ce^{-\lambda x}$ with
parameter $\lambda = 0.230$.  

\subsection{Geographic distributions of origins and destinations}\label{ref:geodist}
\begin{figure}[htbp]
\centering
  \subfloat[2010.10 (Origins)] {
    \includegraphics[scale=.28]{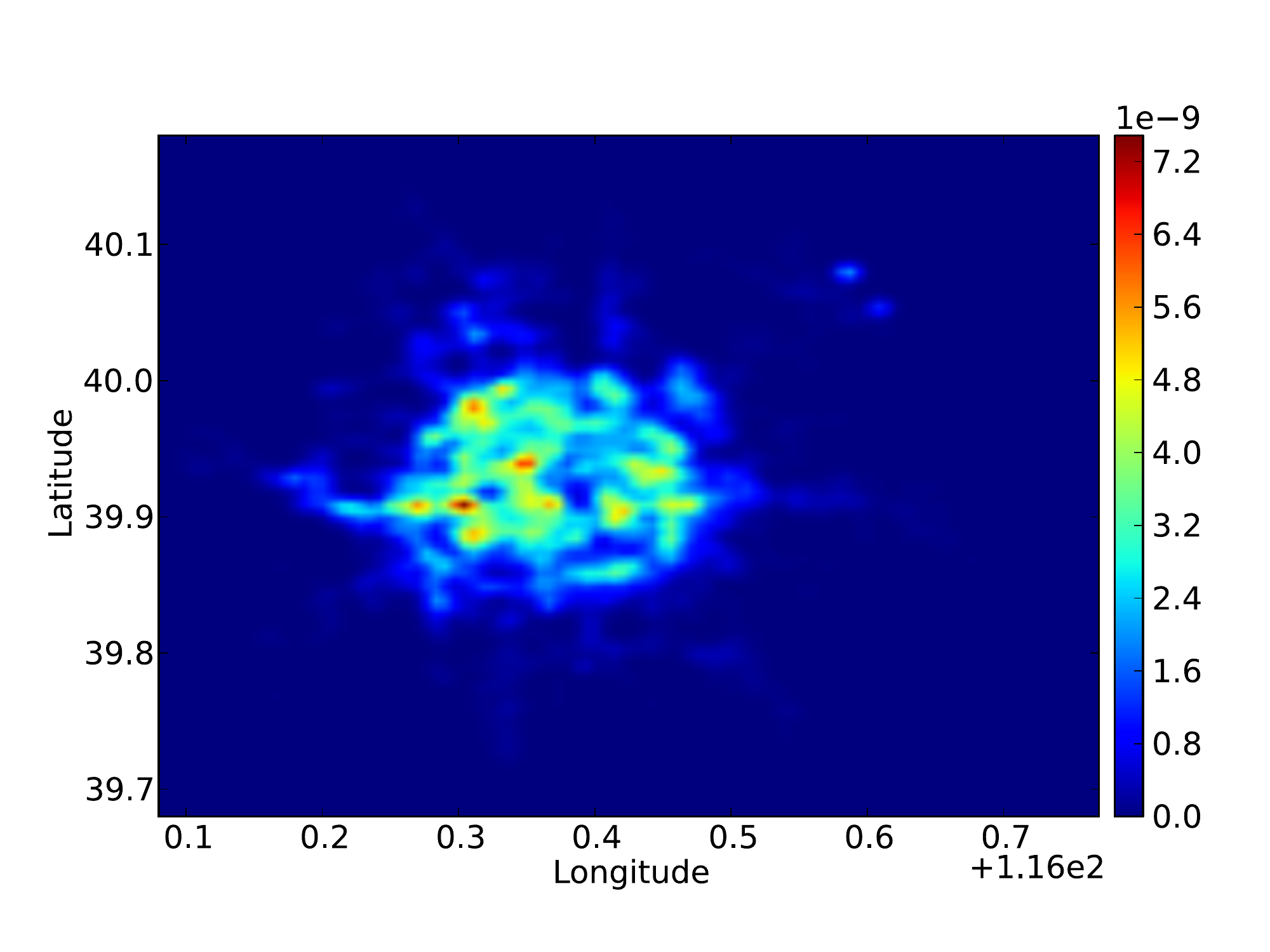}
  }
  \subfloat[2010.11 (Origins)] {
     \includegraphics[scale=.28]{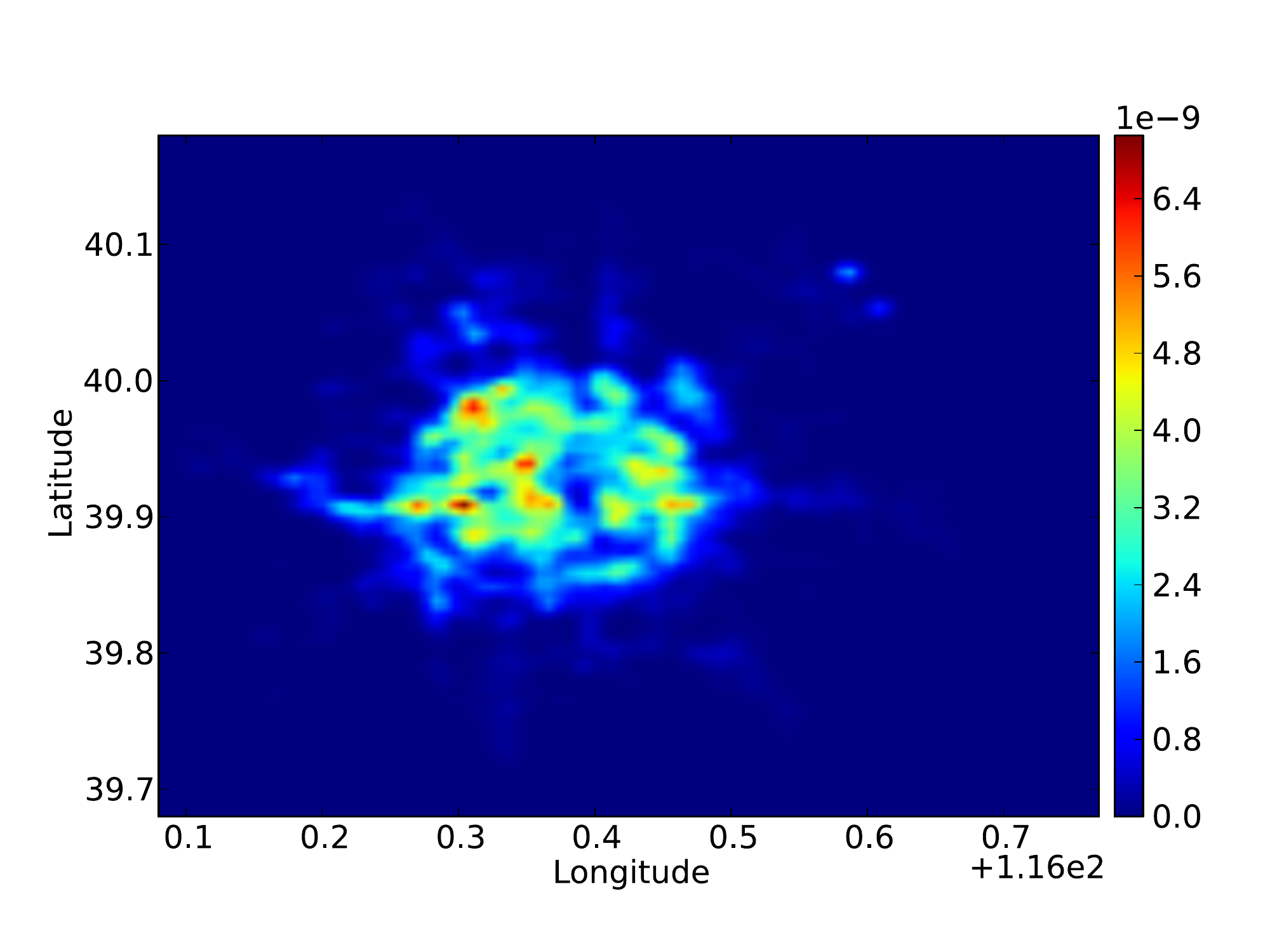}
  }
  \subfloat[2010.12 (Origins)] {
     \includegraphics[scale=.28]{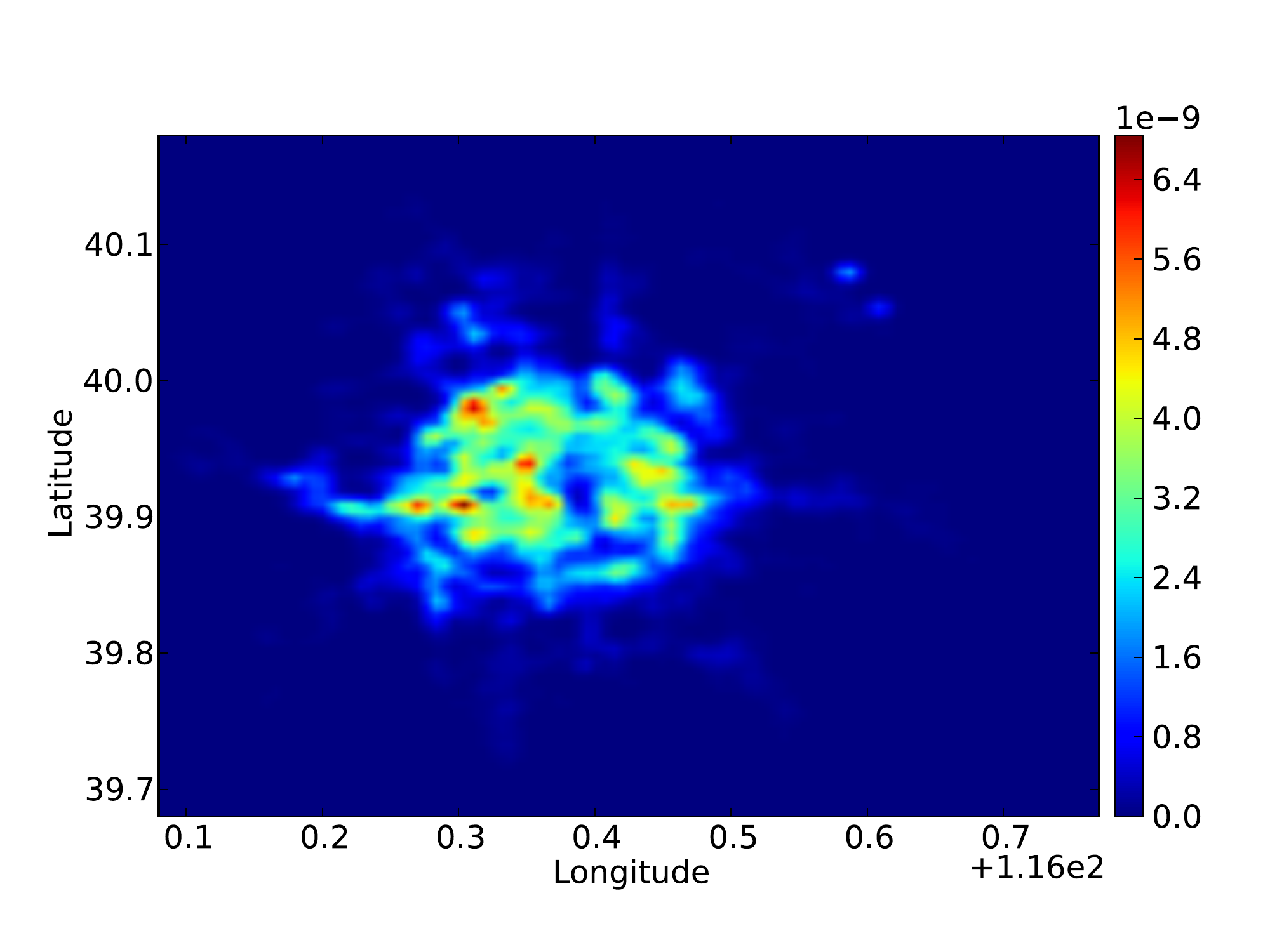}
  }\\
  \subfloat[2010.10 (Destinations)] {
     \includegraphics[scale=.28]{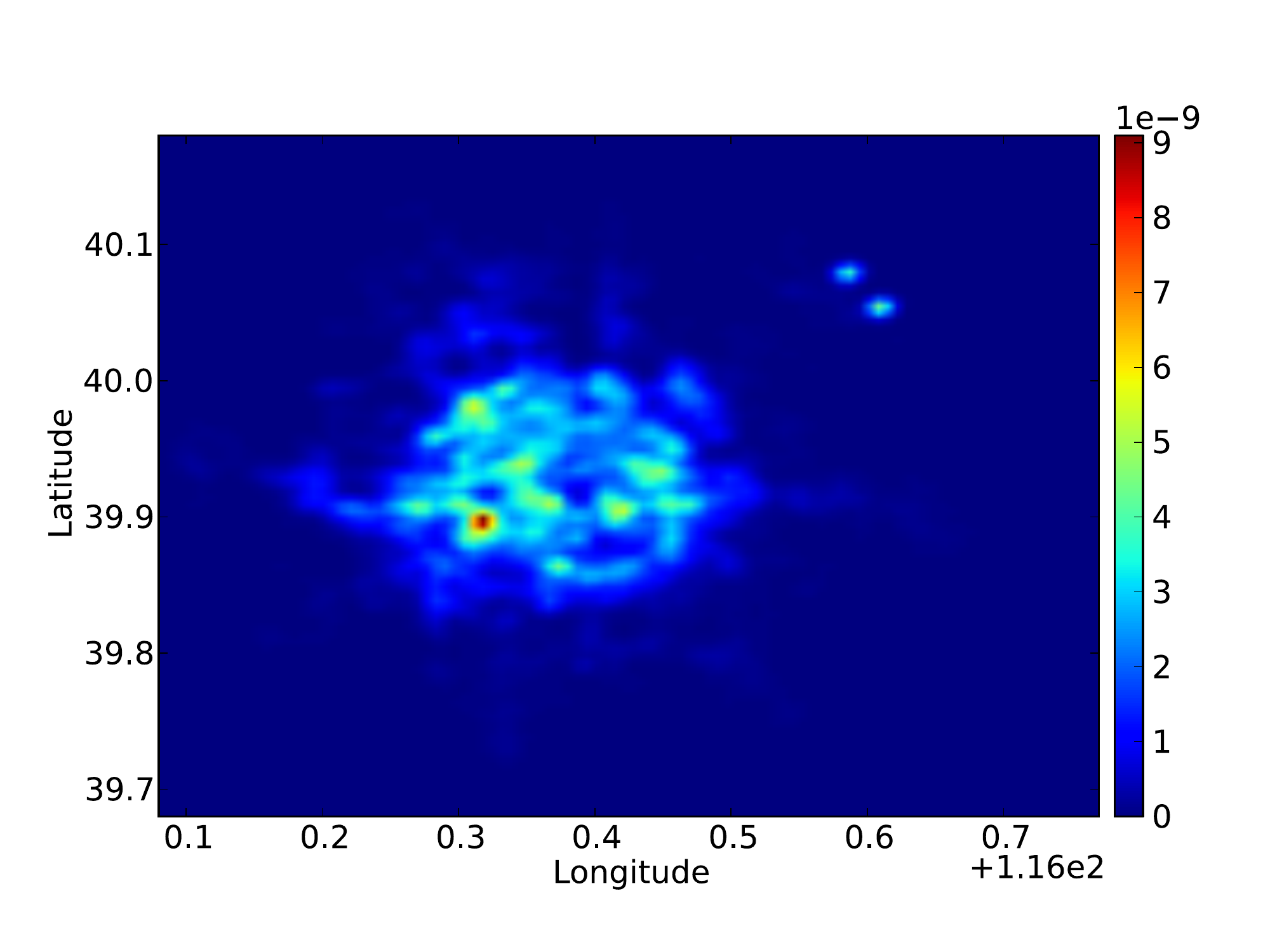}
  }
  \subfloat[2010.11 (Destinations)] {
     \includegraphics[scale=.28]{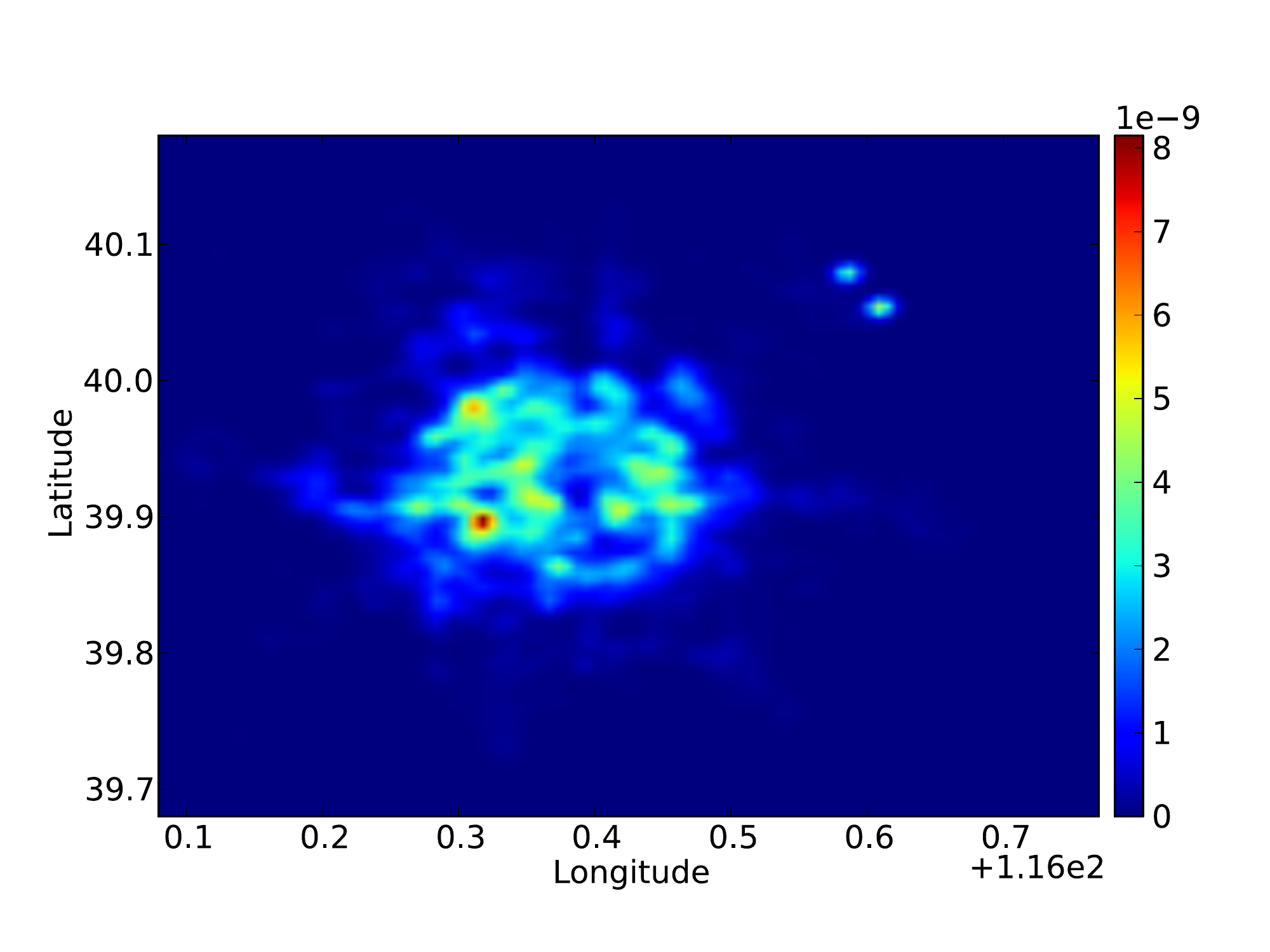}
  }
  \subfloat[2010.12 (Destinations)] {
     \includegraphics[scale=.28]{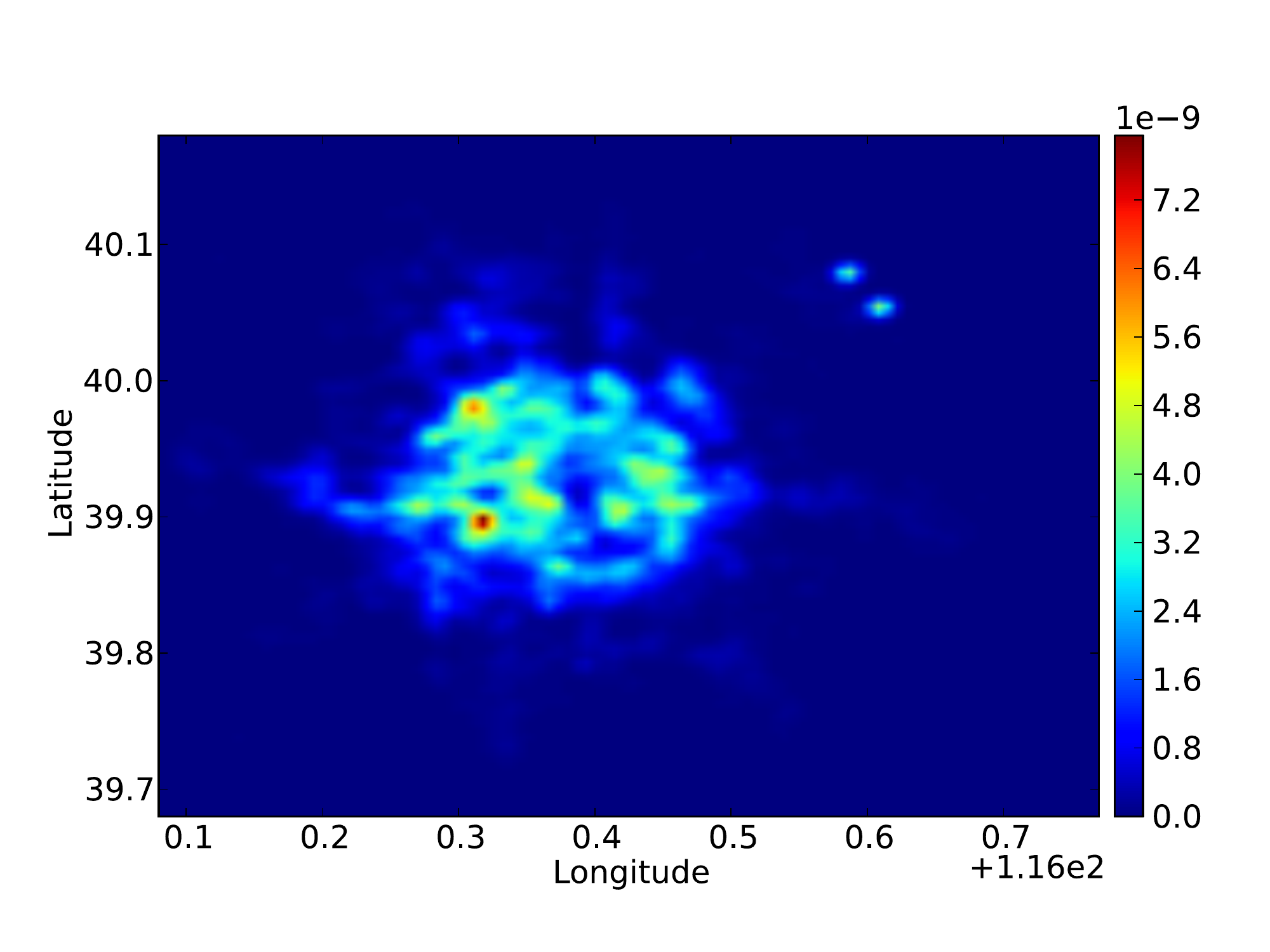}
  }
  \caption{Gaussian kernel density estimations of origins and
  destinations for three months.}
  \label{fig:kde}
\end{figure}

Considering pick-up and drop-off points of human trajectories
respectively, the probability density distributions of them for three
different months are obtained by Gaussian kernel density estimation
(GKDE) method, which are visualized in Fig. \ref{fig:kde}. From the
figure, it can be seen that density maps of origins/destinations for
the three months have similar hot spots and these hot spots, such as
Zhongguancun, Xidan, Beijing West Railway Station and so on, accord with our
intuition very well.

In order to quantify the similarities among geographic distributions of origins
and destinations for the three months, we assign the probability for
each cell as the fraction of origins/destinations falling into the grid
cell. Actually, the probability distribution defined reflect spatial
distribution patterns of human traveling intensity directly. 
After calculating the discrete probability distributions of
origins/destinations for the three months, the similarity between
distributions can be measured by a 
cosine value. To be specific, assuming two discrete probability
distributions $\{p_i\}$ and $\{q_i\}$ ($i=1,\ldots, N$), the
similarity $Sim_{cos}$ between them is defined by
$$Sim_{cos} = \frac{\sum_{i=1}^{N}{p_i q_i}}{\sqrt{\sum_{i=1}^{N}{p_i^2}}\sqrt{\sum_{i=1}^{N}{q_i^2}}}.$$
So, the similarity is assigned to a value between 0 and 1. 
The nearer the value approaches one, the more similar the two
distributions are. The similarities between all distributions of
origins/destinations for the three months are shown in Table \ref{tbl:sim}.
It is noticed that most values of similarity
are larger than 0.95, which indicates that distributions, whether
between origins and destinations or between different months, all
resemble each other and follow similar patterns. In addition, 
it demonstrates that collective travel behaviors in different regions of
urban areas are stable over time.

\begin{table}[htbp]%\tiny
\caption{The cosine similarities among distributions of origins/destinations
  for three months} \label{tbl:sim}
\begin{center}
\begin{tabular}{@{\vrule height 10.5pt depth4pt
      width0pt}ccccccc}
\hline
            &  201010(O)   &  201011(O)       &  201012(O)       &
            201010(D)       &  201011(D)       &  201012(D)\\
\hline
 201010(O)  & 1.0       &  0.996        &  0.995  &  0.951  &  0.957  &  0.961  \\
 201011(O)  & 0.996    & 1.0            & 0.999  &  0.945  &  0.958  &  0.963  \\
 201012(O)  &  0.995   &  0.999       &  1.0     &  0.943  &  0.956  &  0.963  \\
 201010(D)  &  0.951   &  0.945       &  0.943 & 1.0      & 0.996  &  0.994  \\
 201011(D)  &  0.957   &  0.958       &  0.956  & 0.996  &  1.0      &  0.999  \\
 201012(D)  &  0.961   &  0.963       &  0.963  & 0.994  & 0.999    &  1.0      \\
\hline
\end{tabular}
\end{center}
\end{table}

\begin{figure}[htbp] 
  \subfloat[Taxis(2010.10.1--2010.12.31)] {
    \includegraphics[scale=.4]{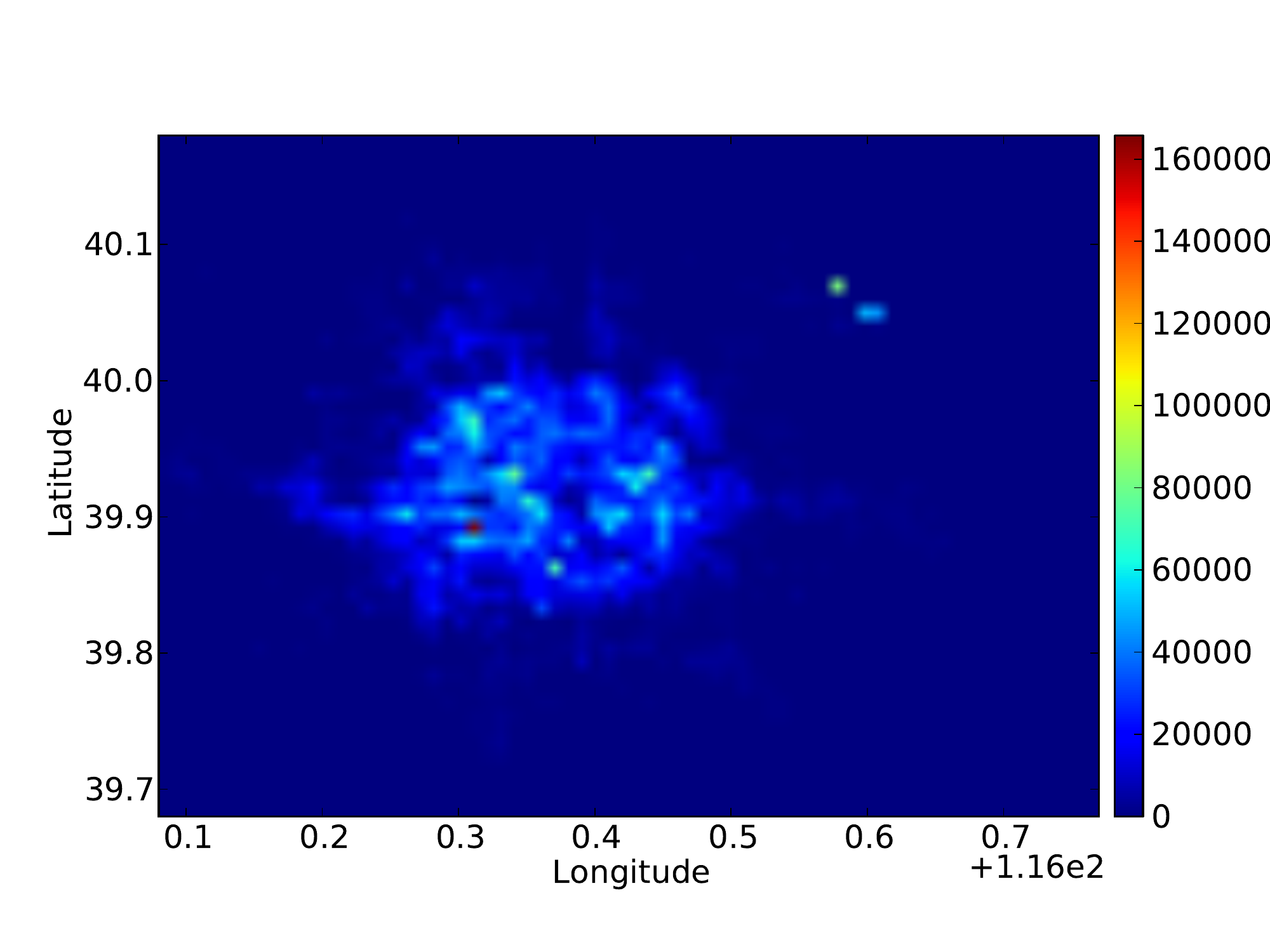}
    \label{fig:taxi_heatmap}
  }
  \subfloat[Sina Weibo(2012.10.8--2012.11.4)] {
    \includegraphics[scale=.4]{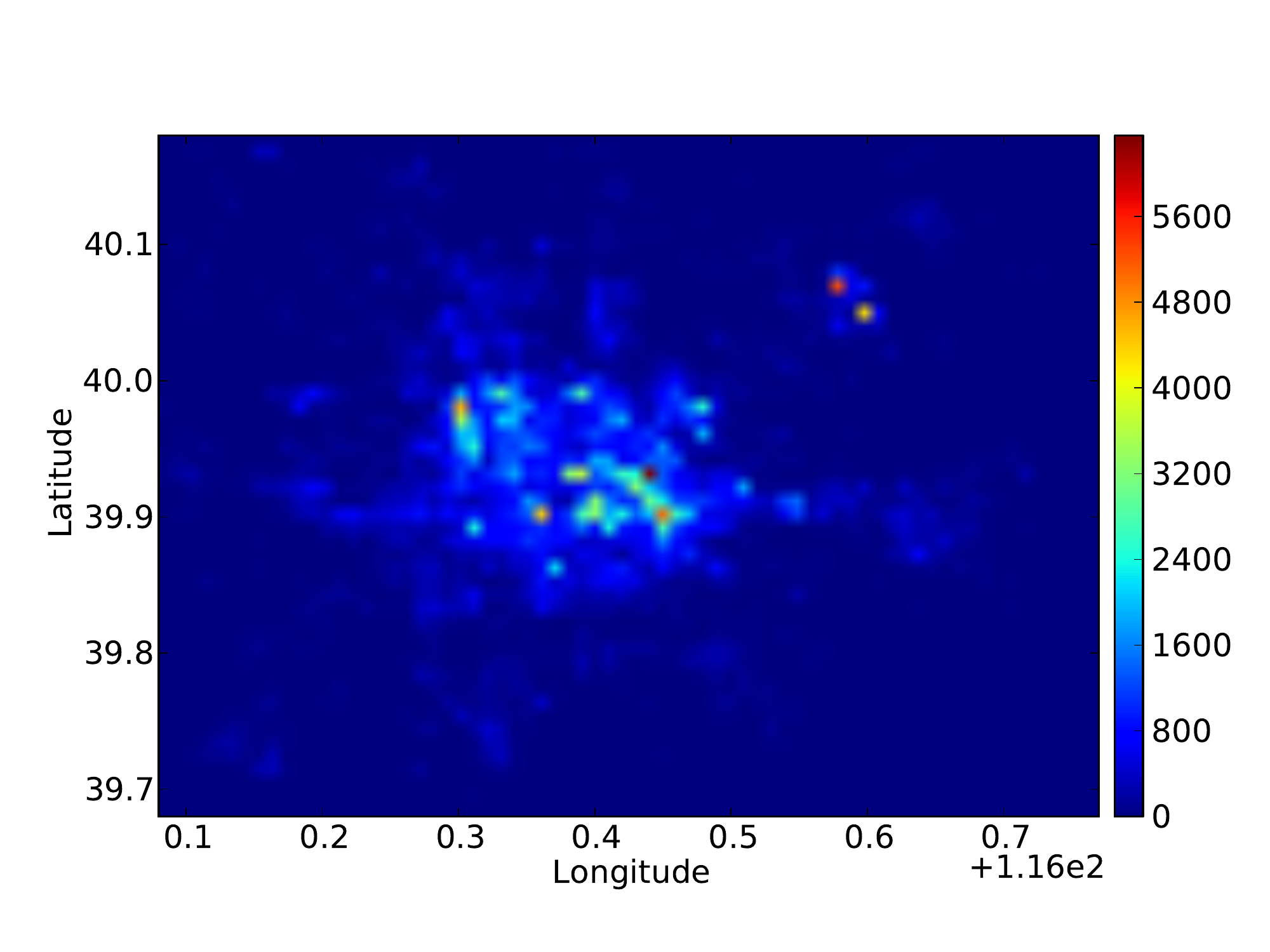}
    \label{fig:weibo_heatmap}
  }
  \caption{The geographic distributions for the two datasets.}
  \label{fig:heatmap_datasets}
\end{figure}

But it must be noted that geographic distributions of
origins/destinations considered here are only based on human travels by taxis. It is
not clear whether there is obvious bias with all intra-urban movement
by different kinds of transport, including private cars, buses, subways,
and taxis. Thus, the geo-tagged Sina Weibo data during 4 weeks in 2012
are collected for comparison (The data description is given in
appendix \ref{app:weibo}).
From the geographic locations of posts, movements of Weibo users in
Beijing are observed. Like the dataset of taxis, it also can
characterize human trails in different geographic regions, but
is independent of means of transportation. Then the geographic distributions of
passengers' destinations and geo-tagged microblogs' locations are
compared, which is illustrated in Fig. \ref{fig:heatmap_datasets}.
From the graph, it can be seen that hot spots reflected by the two
datasets are similar especially within the Fourth Ring Road of the
city. The main difference between them is that the spectrum of human
travel in Sina Weibo dataset is larger than one in taxis dataset. It is
mainly because the urban planning of Beijing has been extending outward for the two
years. And the cosine similarity between these two distributions is 0.829,
which is high and validates our impressions further. 
Because of the similar distribution patterns of destinations and
microblogs' locations described by the two different datasets, it can
be conjectured that means of transportation has only a very small
impact on the geographic distribution of origins/destinations of human
travel in urban areas.

Note that some studies are able to infer land use and regional
functions successfully through analyzing
spatiotemporal variation of human movement captured from passengers
by taxis \cite{Zheng2011} or cellphone users \cite{Toole2012}. In
fact, because geographic distribution of origins/destinations is
influenced by demands for mobility, these all demonstrate that
the distribution patterns of traveling demands or intensity are
mainly related to inherent properties of the city, such as urban
planning, regional functions, population density and so on, rather
than means of transport.

\section{Modeling intra-urban human mobility}
In the section, it is aimed to explore urban flows using tracks of
passengers by taxis. Because the urban area has been divided into grid
cells, we try to model and predict traffic flows between these grid
cells.
\subsection{Radiation model}
Recently, the radiation model \cite{Simini2012a}, which is parameter-free and only
requires information of population distribution, was proposed to 
characterize mobility patterns between cities or countries. It
overcomes some shortcomings of the gravity model and predicts traffic
flows more precisely. So it is wished to verify
whether the radiation model can be utilized to simulate
collective human travel in urban areas of the city. But in urban areas, it is difficult to obtain
population distribution directly because of high mobility of people. Instead we use the
distribution of destinations in the simulation. It is more
reasonable because the number of destinations in a
grid cell describes human travel intensity in the region.

\begin{figure}[htbp] 
  \subfloat[Prediction(radiation model)] {
    \includegraphics[scale=.4]{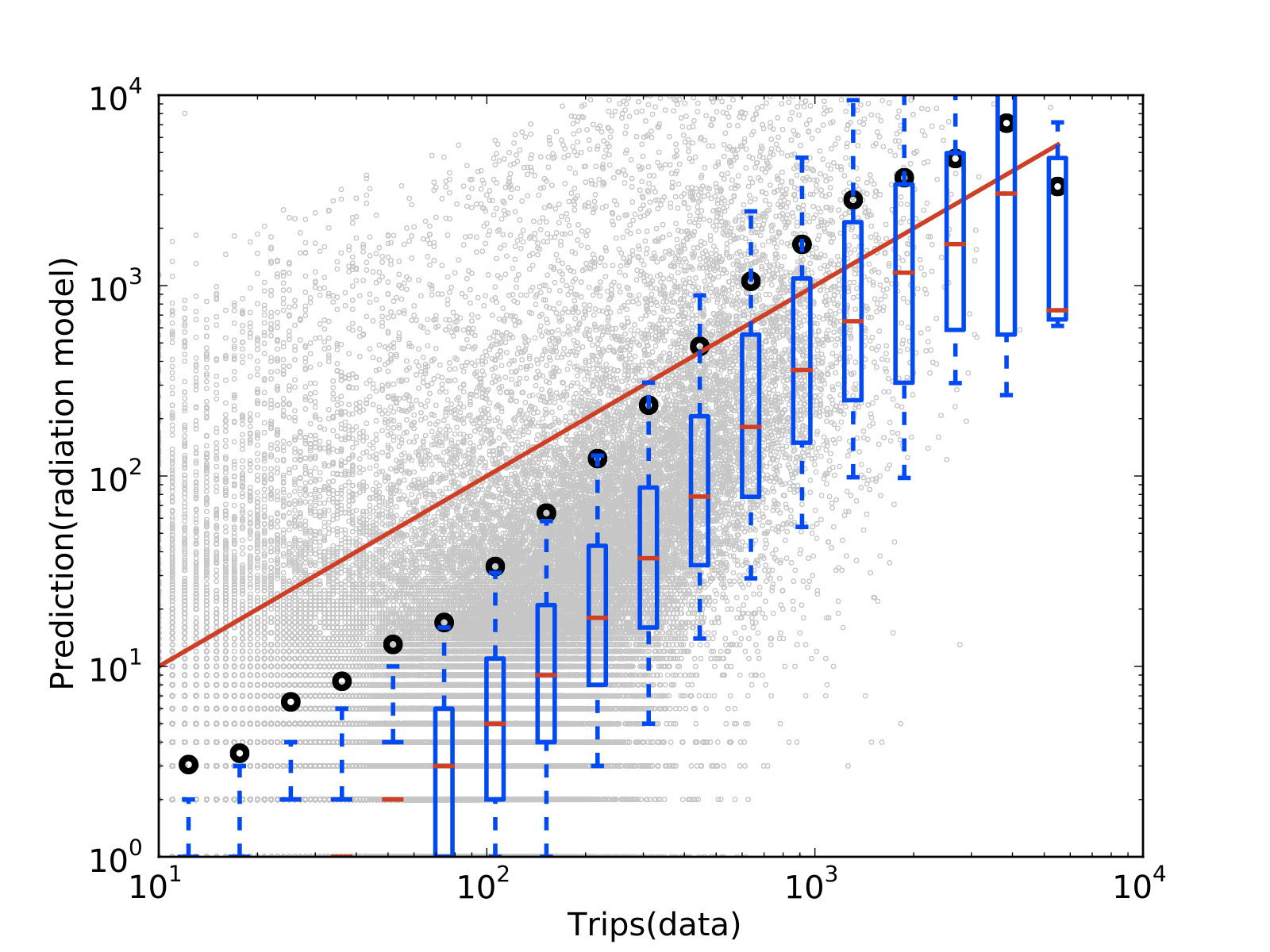}
    \label{fig:sim_radia_pre}
  }
  \subfloat[Probability Distribution of trip length] {
    \includegraphics[scale=.4]{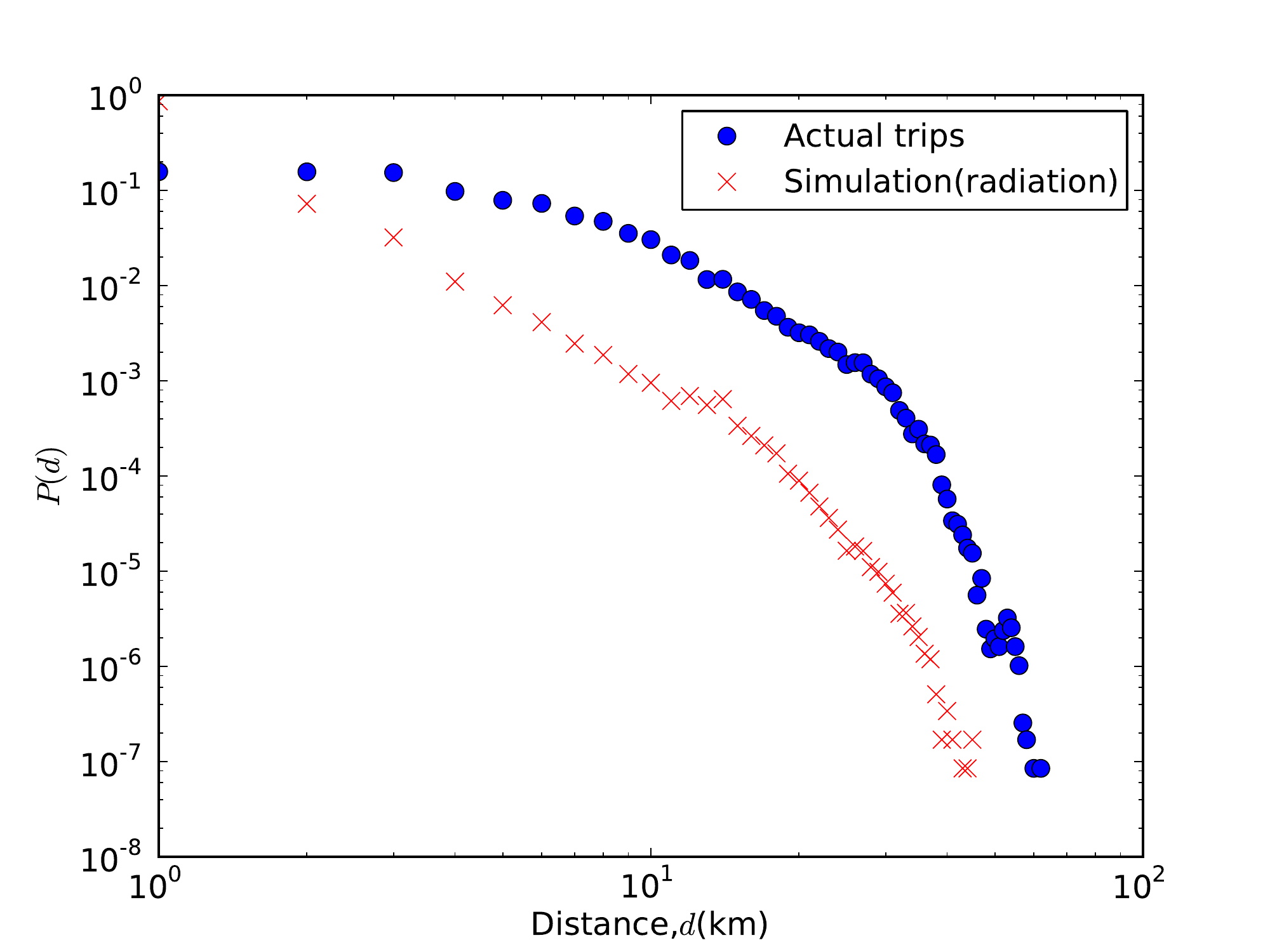}
    \label{fig:sim_radia_disp}
  }
  \caption{Simulation by the radiation model. (a) Prediction of
    traffic flows between grid cells. Grey points stand for the
    relationship between actual and predicted flux of each
    pair of grid cells. The red line $y=x$ stands for the actual
    values equal with predicted values. The black circles denote mean
    values of prediction in the bins. The ends of whisker represent the 9th and 91st
    percentile in the bin. (b) Probability distributions of actual and predicted
    trip length.}
  \label{fig:sim_radiation}
\end{figure}

The results of simulation by the radiation model is shown in
Fig. \ref{fig:sim_radiation}. From the figure, it seems that the predicted
flux has a large deviation from the actual ones and the model
underestimates the probability of trips with distances larger than 1km.
There are two possible reasons: one is that the destinations
distribution may have some bias with actual population distribution;
the other is that, unlike trips between countries or cities, the
population distribution may be only  
one of factors to influence human movement because people often move
frequently for various purposes in urban areas. Therefore, it is necessary to
consider a new model to understand intra-urban human mobility patterns.

\subsection{Our model}
As for trips occurred during a period of time, it can be calculated that
how many of them had left from or arrived at each cell. So the
probability that people leave from/arrive at each cell is defined as follows
\begin{eqnarray*}
P_{O}(i) &=& \frac{\#\text{ of trips leaving from the cell }i}{\#\text{
of trips}},\\
P_{D}(i) &=& \frac{\#\text{ of trips arriving at the cell }i}{\#\text{
of trips}}, i=\{1,2,\ldots,N\}.
\end{eqnarray*}
Actually, $P_{O}$ and $P_{D}$ correspond to the distributions of
origins and destinations separately.
As demonstrated in subsection \ref{ref:geodist}, it must be noted that $P_{O}$ and $P_{D}$ only depend on
population distribution, urban planning (land-use and transportation
planning) and other environmental factors. The cells having large $P_{O}$ or $P_{D}$ often lie in
prosperous commercial/entertainment regions, developed residential
areas, transport hubs and so on.

Intuitively, the probability of a trip's occurrence has positive correlations
with $P_{O}$ of origin cell and
$P_{D}$ of destination cell, but has a negative correlation with the
Euclidean distance between the two cells.
Hence, in our model the probability of a trip reaching cell $j$,
conditioned on starting from cell $i$, is defined as follows
$$P(j|i) \propto \frac{P_D(j)}{f(d_{ij})}$$
where $f(d)$ is a function of distance between cells. In the paper,
two frequently used forms of $f(d)$ are given by
\begin{eqnarray*}
Power-law&:&f(d) = d^\sigma\\
Exponential&:&f(d) = e^{\lambda d}
\end{eqnarray*}
where $\sigma$ and $\lambda$ are parameters whose values rely on the
specific system and reflect the effect of distance on human
travel. So the happening probability of a trip from cell $i$ to $j$ can be
derived as
\begin{eqnarray}\label{formula:prob}
P(i\to j) &=& P_{O}(i)P(j|i) \nonumber \\
       &=&
       P_{O}(i)\frac{P_{D}(j)/f(d_{ij})}{\sum_{k\neq i}{P_{D}(k)/f(d_{ik})}}.
\end{eqnarray}
And the expected number of trips from cell $i$ to $j$ can be obtained
\begin{eqnarray}\label{formula:pre}
\langle T_{ij}\rangle &=& TP(i\to j) \nonumber \\
       &=&
       \frac{T}{\sum_{k\neq
           i}{P_{D}(k)/f(d_{ik})}}\frac{P_{O}(i)P_{D}(j)}{f(d_{ij})}
       \nonumber \\
& = & \frac{T}{M(i)}\frac{P_{O}(i)P_{D}(j)}{f(d_{ij})}.
\end{eqnarray}
where $M(i) = \sum_{k\ne i}{P_{D}(k)/f(d_{ik})}$ and $T$ is the total number of trips.

Note that it can be concluded that $P_O\approx P_D$ because the
cosine similarity between distributions of origins and destinations is
greater than 0.95 in taxis dataset. Furthermore, when considering
human travel during a long time, it is reasonable to assume $P_O =
P_D$ due to round-trip patterns in urban areas. As described in
the gravity model, the number of trips from cell $i$ to $j$ is equal
with the one from cell $j$ to $i$. However, that is not the case in
our model because the values of $M(i)$ and $M(j)$ depend on locations
of $i$ and $j$ respectively, which are often not equal to each other.
Therefore, it is more consistent with our intuition. 

\begin{figure}[htbp] 
  \subfloat[Simulation by our model (power-law)] {
    \includegraphics[scale=.4]{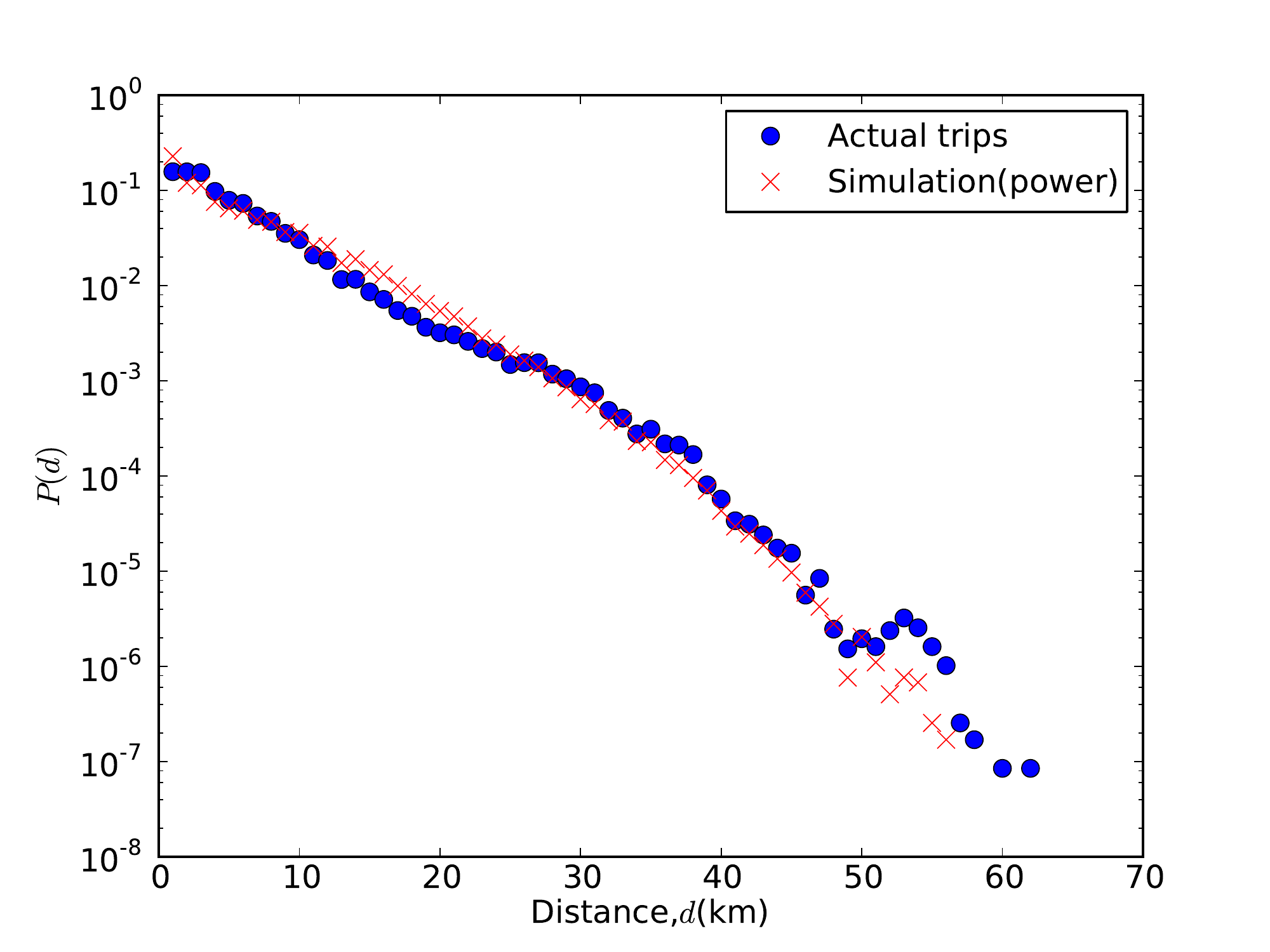}
    \label{fig:sim_power}
  }
  \subfloat[Simulation by our model (exponential)] {
    \includegraphics[scale=.4]{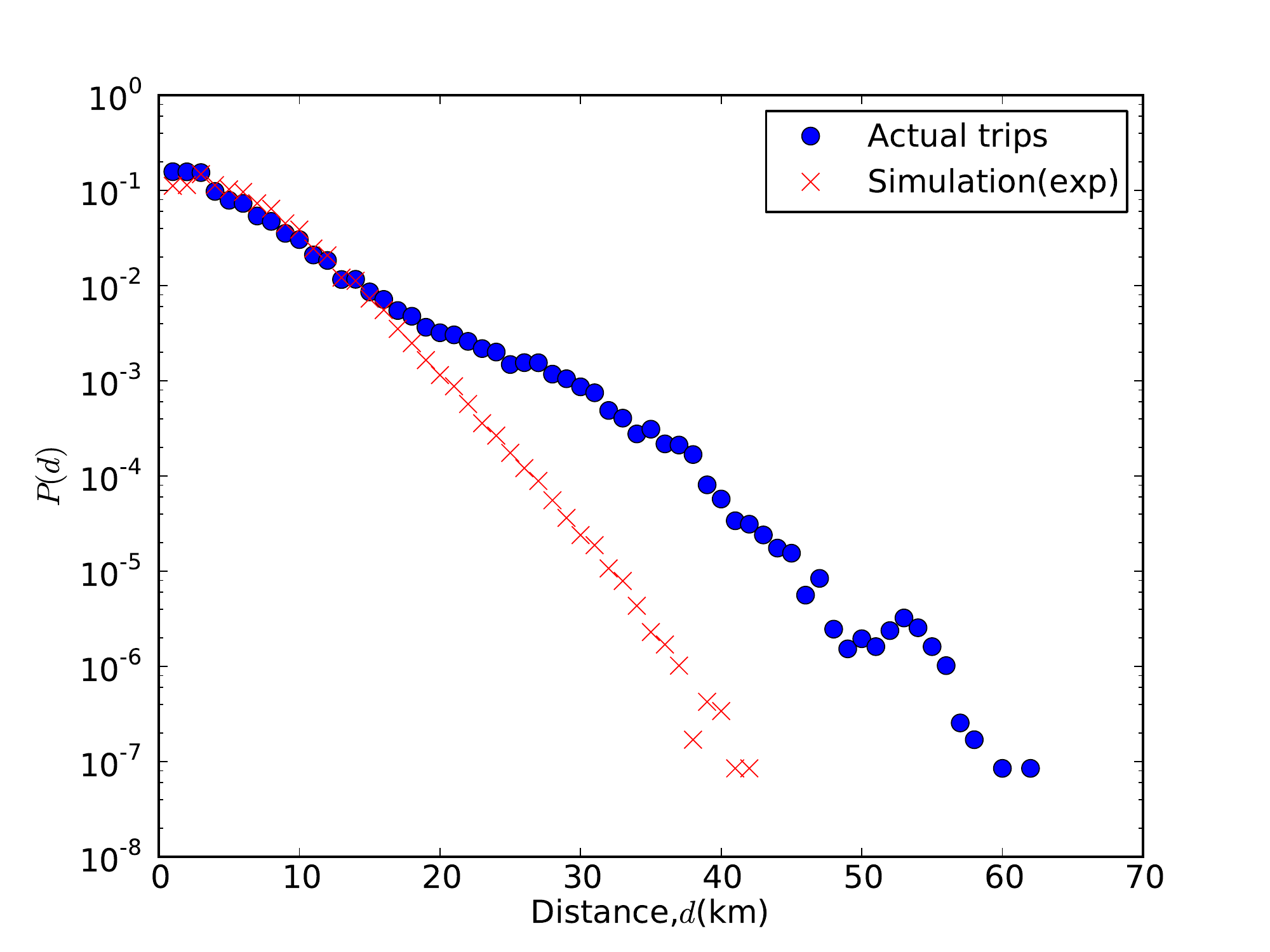}
    \label{fig:sim_exp}
  }\\
  \subfloat[Prediction by our model (power-law)] {
    \includegraphics[scale=.4]{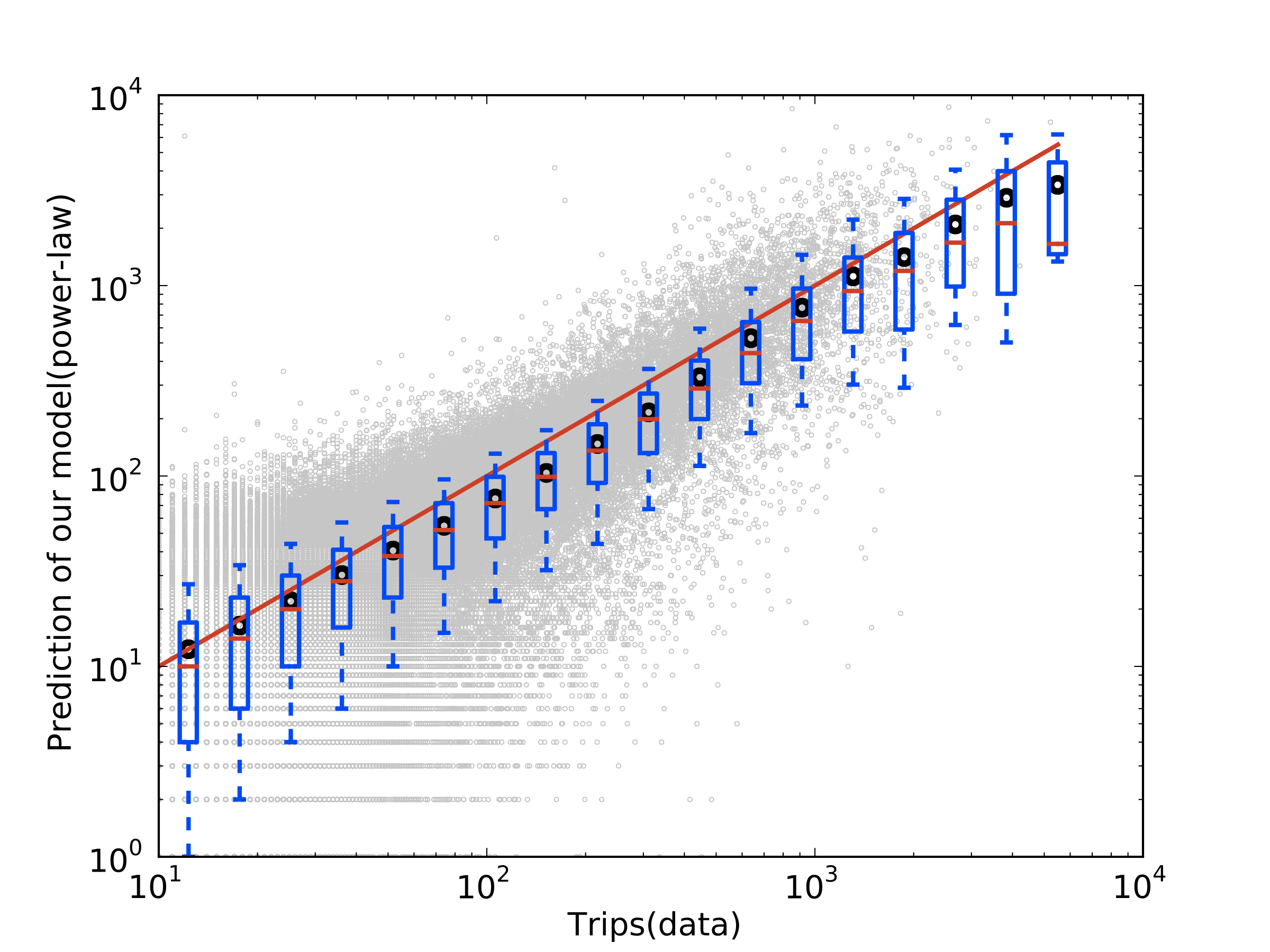}
    \label{fig:pre_power}
  }
  \subfloat[Prediction by our model (exponential)] {
    \includegraphics[scale=.4]{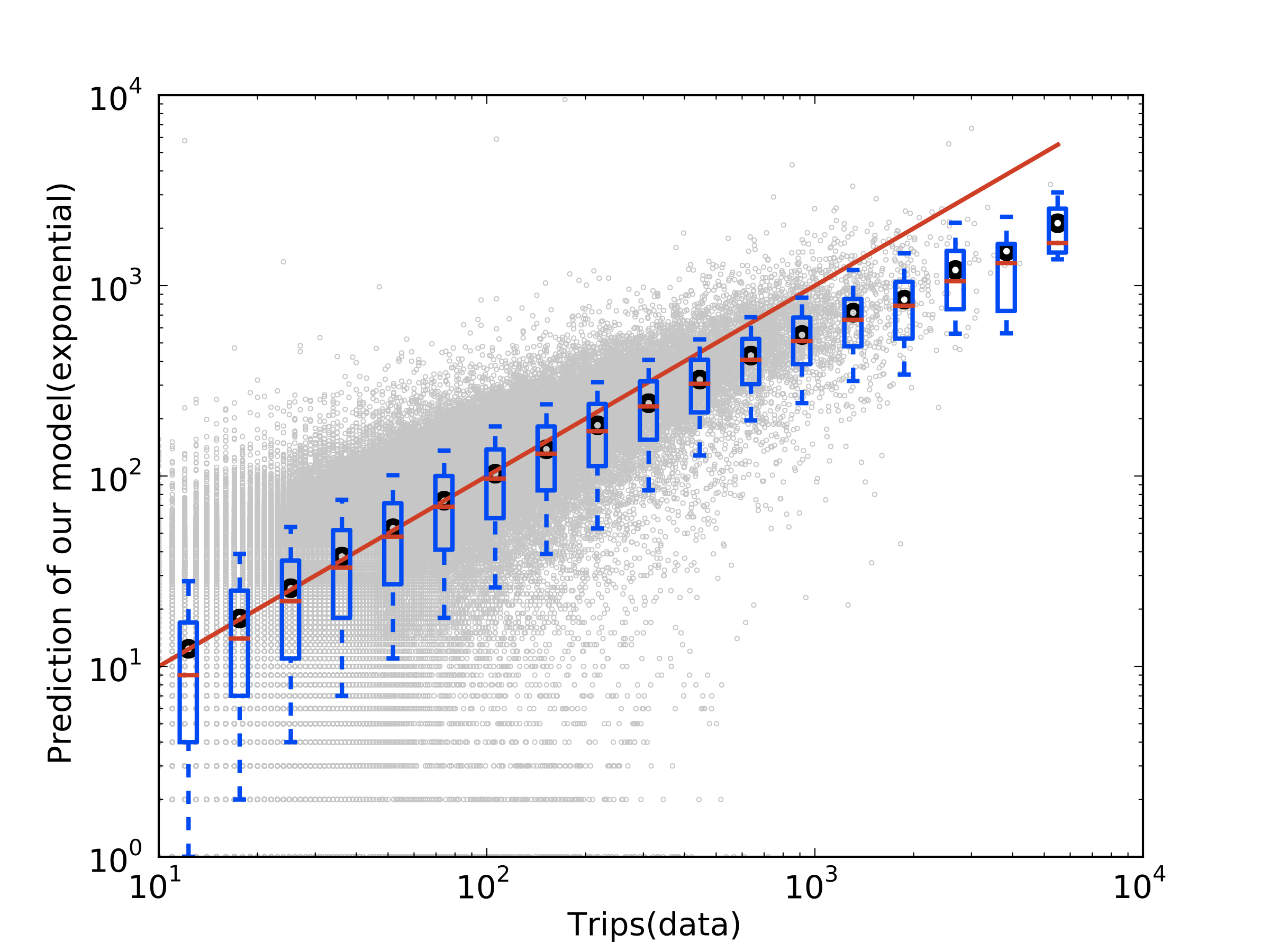}
    \label{fig:pre_exp}
  }
  \caption{Simulations by the two forms of our model. (a-b) Distributions of trip
    length simulated by our model. (c-d) Prediction of traffic flows
    between cells by our model.}
  \label{fig:simulation}
\end{figure}

Subsequently, we apply our model to simulate trips by taxis in urban
areas. The method of Maximum Likelihood Estimation (MLE) is used to evaluate the
parameters in our model (see the appendix \ref{add:mle} for details). As for
the two forms of function $f(d)$, the parameters $\sigma$ and $\lambda$
of our model are calculated as 1.601 and 0.308 respectively. 
The Fig. \ref{fig:sim_power} and \ref{fig:sim_exp}, corresponding to
the two forms of $f(d)$ in our model, both describe comparisons of
distance distributions between simulated trips and actual ones.
It can be seen that the distance distribution of trips predicted by our model with the form
of power-law can accord with the actual ones very well and instead our
model with the form of exponential underestimates the amount of
trips with long distance. Furthermore, actual and predicted traffic
flows between pairs of cells by the two forms of our model are shown in Fig. \ref{fig:pre_power} and
\ref{fig:pre_exp}. From the figures, it can be observed that the red line $y = x$ lies
between the 9th and the 91st percentiles in bins in our model with the
form of power-law, indicating that the model can predict the
number of trips between cells accurately. But our model with the form
of exponential may underestimate traffic flows. In summary, our
model with the form of power-law can 
be treated as an appropriate model to predict traffic flows in urban
areas. Thus, in the following paper, we only use our model with the
form of power-law by default.

\begin{figure}[htbp] 
  \subfloat[Radiation model] {
    \includegraphics[scale=.4]{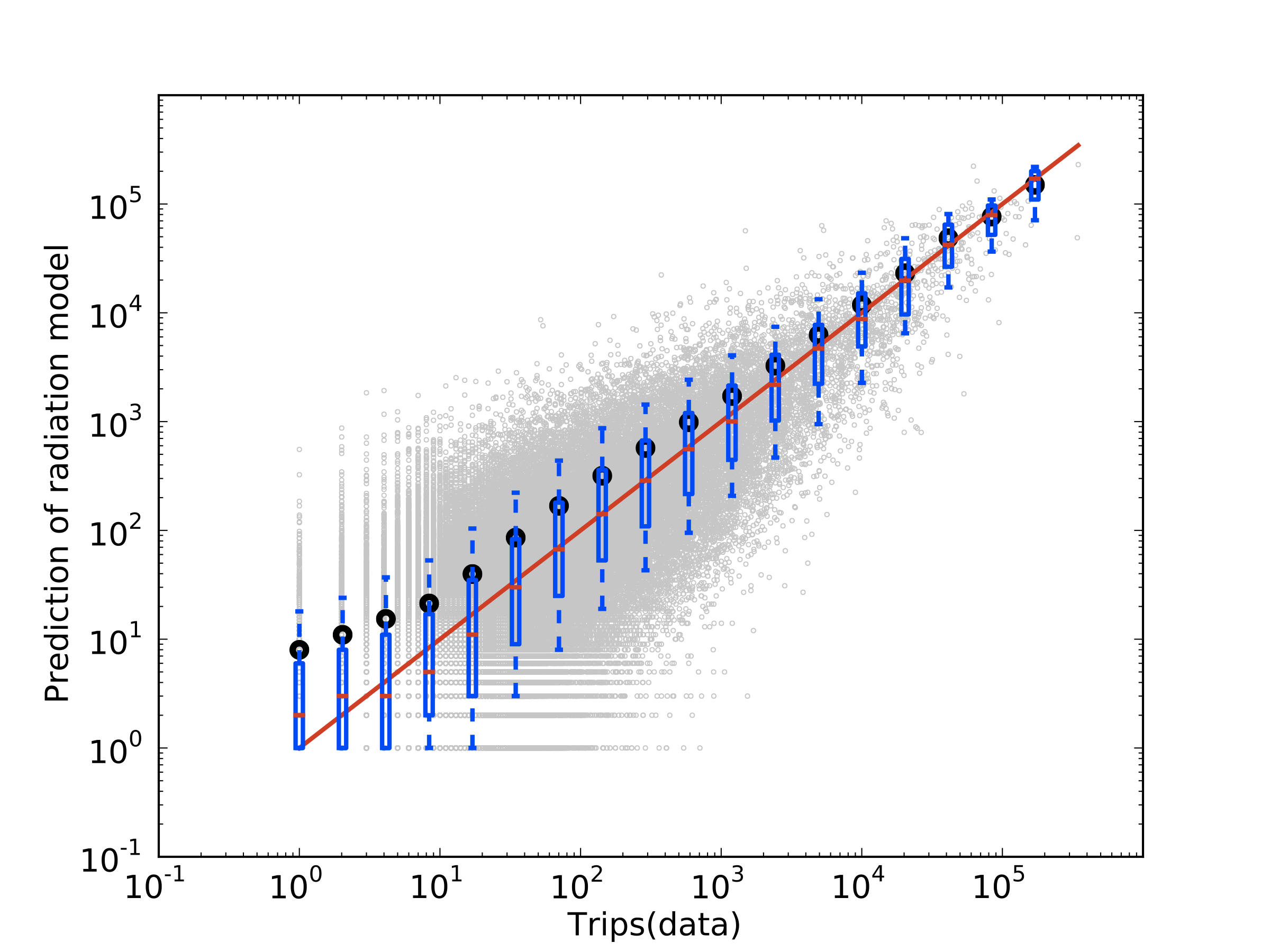}
    \label{fig:us_sim_rad}
  }
  \subfloat[Our model] {
    \includegraphics[scale=.4]{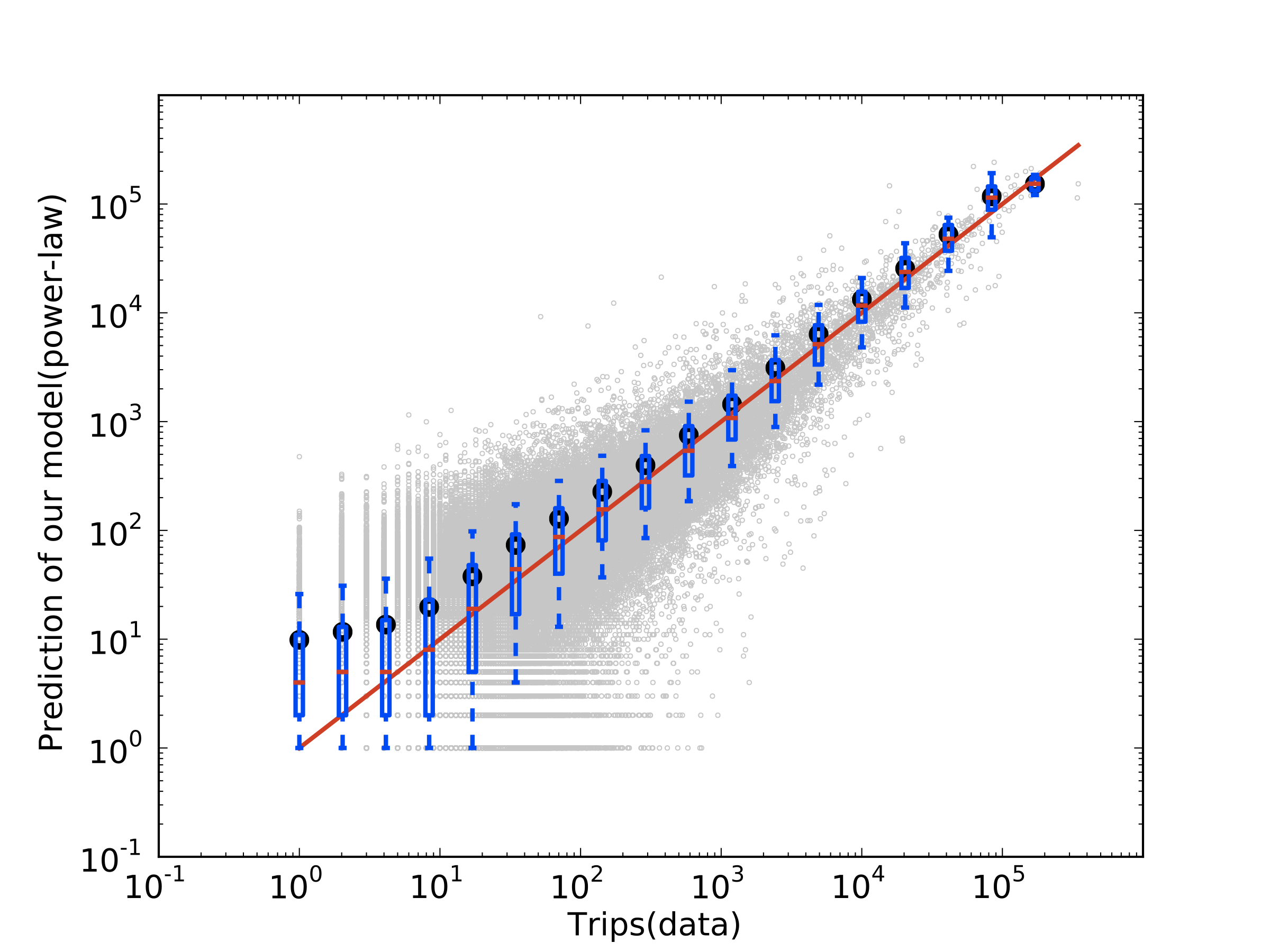}
    \label{fig:us_sim_model}
  }
  \caption{Predictions on US commuting.}
  \label{fig:us_sim}
\end{figure}

As demonstrated before, our model can be suitable to simulate human travels
by taxis in urban areas. Whether is our model also applied to model
collective human movement at large scale?
Here, the dataset of US commuting is used, which described US commuting between
United States countries in 2000 \cite{Simini2012a}.
As for our model, $P_O$ and $P_D$ shown in formula
(\ref{formula:prob}, \ref{formula:pre}) are replaced by the fractions of population
in countries. The function $f(d)$ takes the form of power-law and $d$
is the distance between countries. So by using the method of MLE, it
can be confirmed that the value of $\sigma$ is 3.077.  
In Fig. \ref{fig:us_sim}, the results of predicting the commuting
number between pairs of counties by radiation model and our model are
shown separately. 
%The red line describes the situation about prediction number
%conforming to actual commuting number. 
Both models can predict the mobility patterns very well
because both red lines almost fall between the 9th and 91st
percentiles in each bin. In addition, observed from the gray points in the
figure, the results predicted by our model seem more compact and have
less fluctuations. 

As a result, it is concluded that our model is very flexible and can simulate human
movement not only in urban areas but also in countries.

\section{Analysis of distance distributions}
In our model, $P_O$ and $P_D$, corresponding to distributions of
origins and destinations, are probabilities of individuals'
leaving from and arriving at cells. For a whole day or longer time, both are
almost equal and only depend on inherent properties of the city. So it is assumed that 
$$P_O(i)=P_D(i)=P(i)\quad (i =1,\ldots,N),$$
where $P(\cdot)$ reflects human travel demands in grid cell regions.

\begin{figure}[htbp]
\centering
\includegraphics[scale=.4]{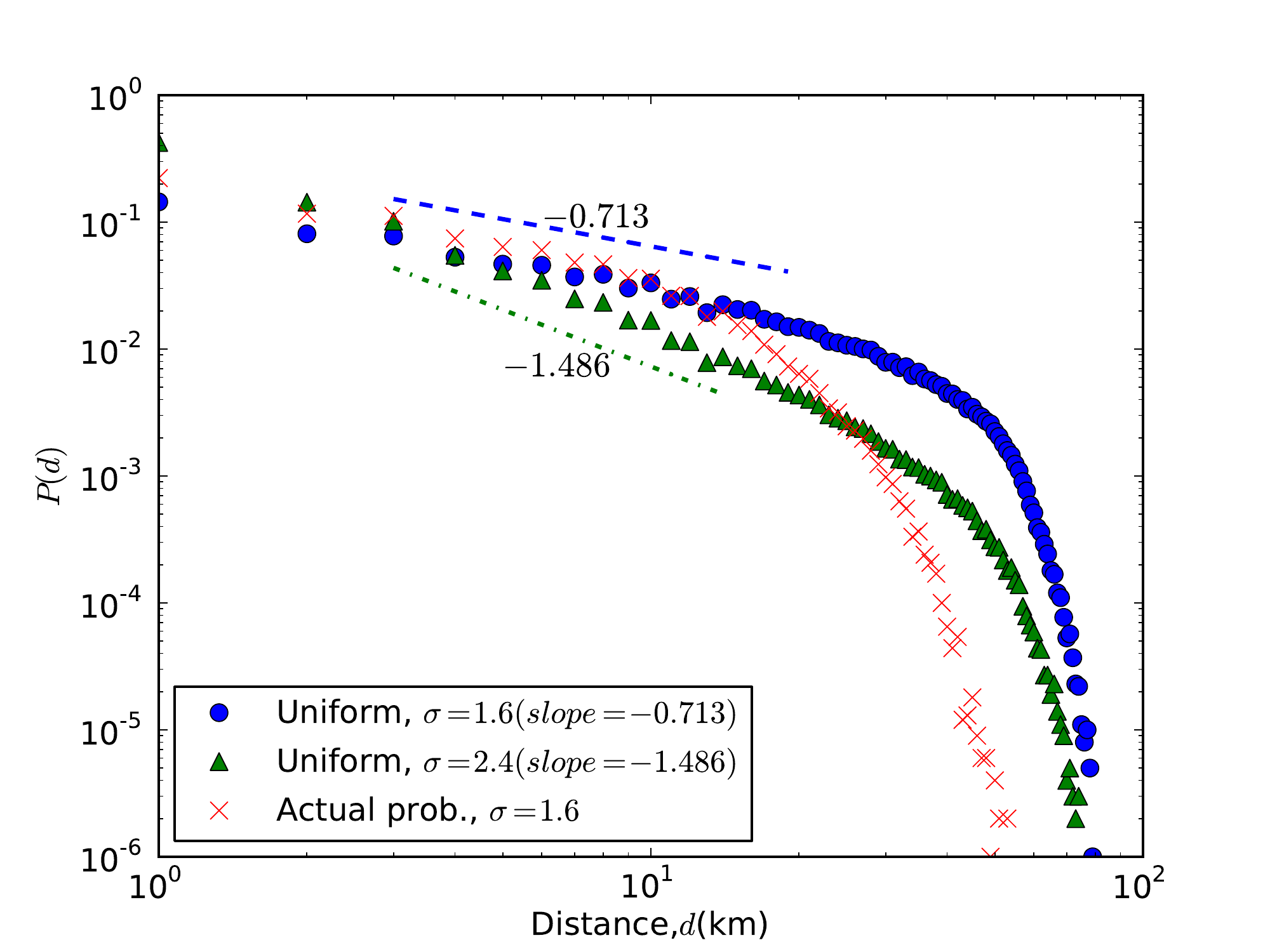}
\caption{The simulations based on uniform distribution of human travel
  demands for different $\sigma$.}
\label{fig:dif_q_uniform}
\end{figure}

Supporting the probability $P(\cdot)$ is uniform, the trips are
simulated based on our model with two different values of $\sigma$
respectively (1.6, 2.4). As shown in Fig. \ref{fig:dif_q_uniform},
it can be seen that the distance distributions of trips in the two simulations accord to power-law
with exponential cutoff very well. Actually, the exponential decay in the tail
is caused by the geographic limits. The exponents of power-law in the
two distributions are -0.713 for $\sigma=1.6$ (blue circles) and -1.486
for $\sigma=2.4$ (green triangles), which approach to the theoretical
results as demonstrated in appendix \ref{add:uniform} and will become more and
more close to it as increasing the number of cells $N$. At the same time, the
simulation according to actual distribution of travel demands in urban area of
Beijing is also shown in the graph (red crosses), which is very close
to the distribution of actual trip length. Compared with the
simulations of uniform distribution, the travel distances decrease more
rapidly even though the same value of $\sigma$.

\begin{figure}[htbp]
\subfloat[Rearrangements] {
\includegraphics[scale=.4]{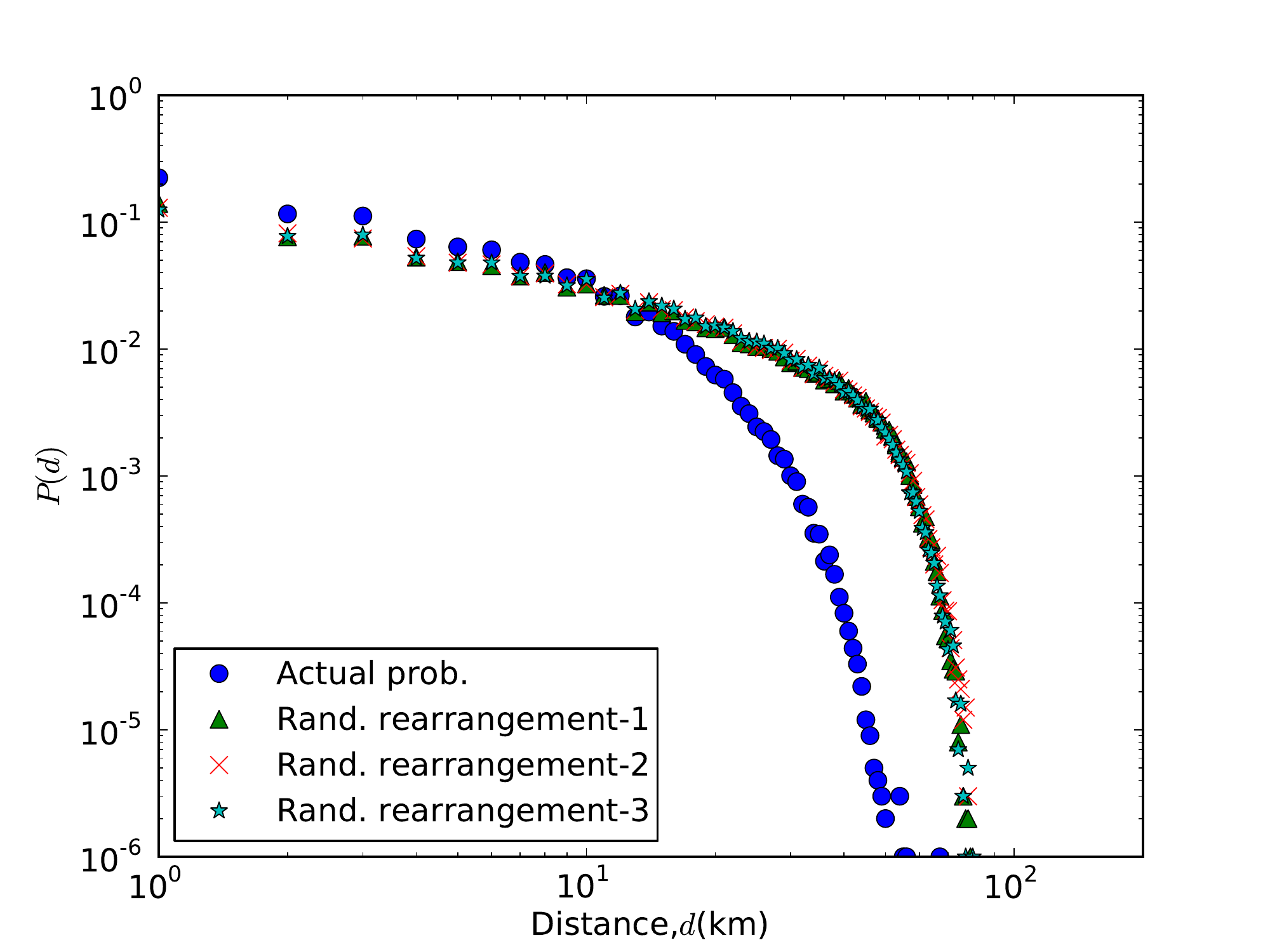}
\label{fig:act_bj_shuffle}
}
\subfloat[Comparison] {
\includegraphics[scale=.4]{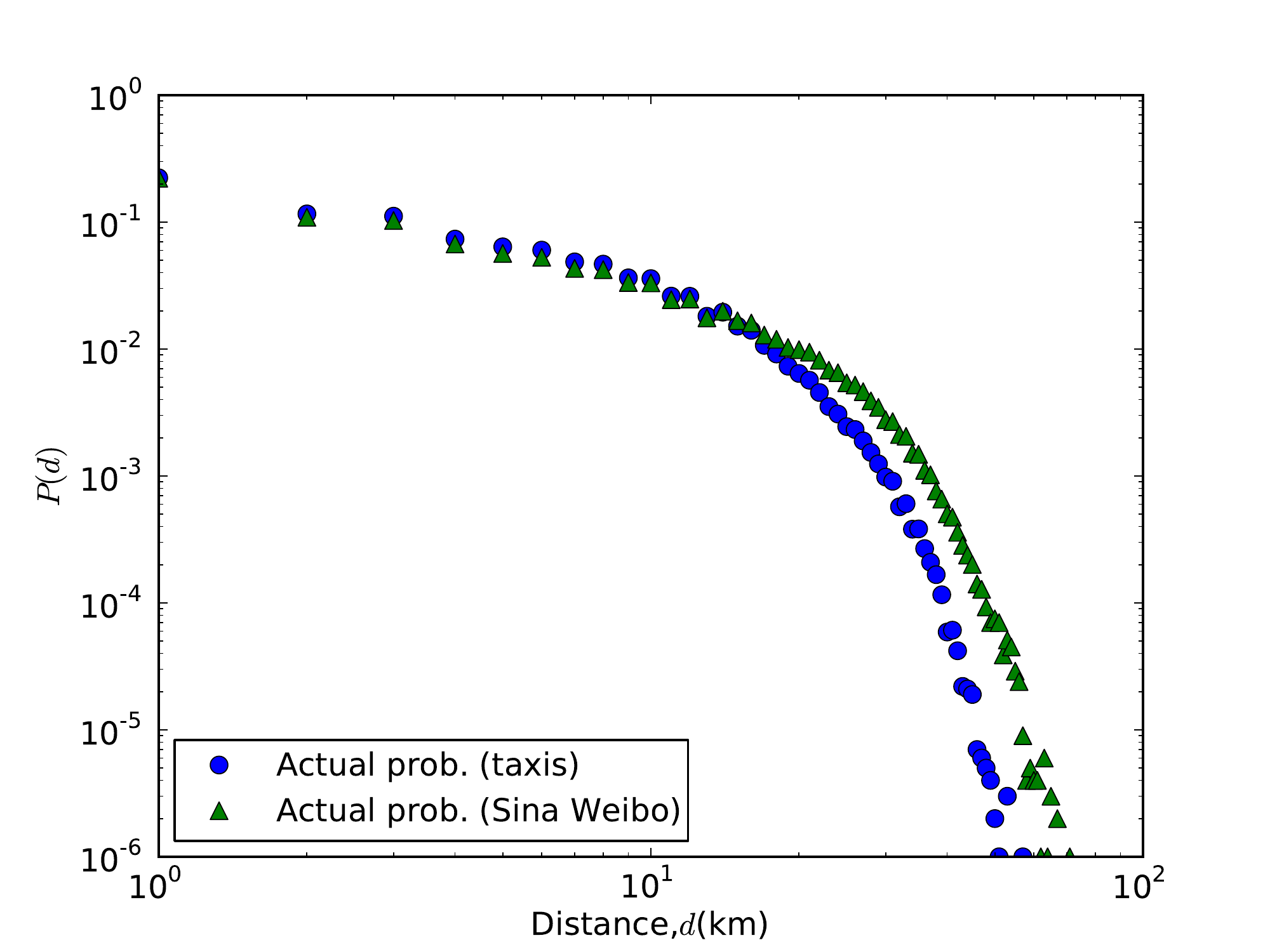}
\label{fig:taxi_weibo_dist}
}
\caption{The simulations on different distributions. (a) Randomized
  rearrangements of actual probability distribution. (b) Comparison
  distributions of trip length based on actual probability
  distribution of taxis and Sina Weibo.}
\end{figure}

Then, considering the probability distribution of $P(\cdot)$ is the same as
the actual one in Beijing, we only rearrange the probability values of grid
cells randomly. As shown in Fig. \ref{fig:act_bj_shuffle}, the travels
simulated on grid cells with randomized permuting probability values accord with
power-laws in the heads and decay more slowly, compared with
ones simulated on grid cells with actual travel demands. And in
Fig. \ref{fig:taxi_weibo_dist}, the two simulations based on actual travel
demands reflected by taxis dataset and Sina Weibo dataset are shown respectively.
It is illustrated that both distributions have similar trends in the 
head, but the tail of trips reflected by taxis dataset decreases more sharply. It
is reasonable because, as described before, both geographic distributions of
travel demands are similar near urban centers and the urban planning of Beijing expands
during the two years.

In conclusion, it suggests that not only the probability distribution of human
travel demands, but also the layout of them is the fundamental element to
account for collective human travel and determine the distribution of trip
lengths. And they only depend on urban planning, population distribution and
other properties of the city directly.

\begin{figure}[htbp]
\subfloat[Taxis] {
\includegraphics[scale=.4]{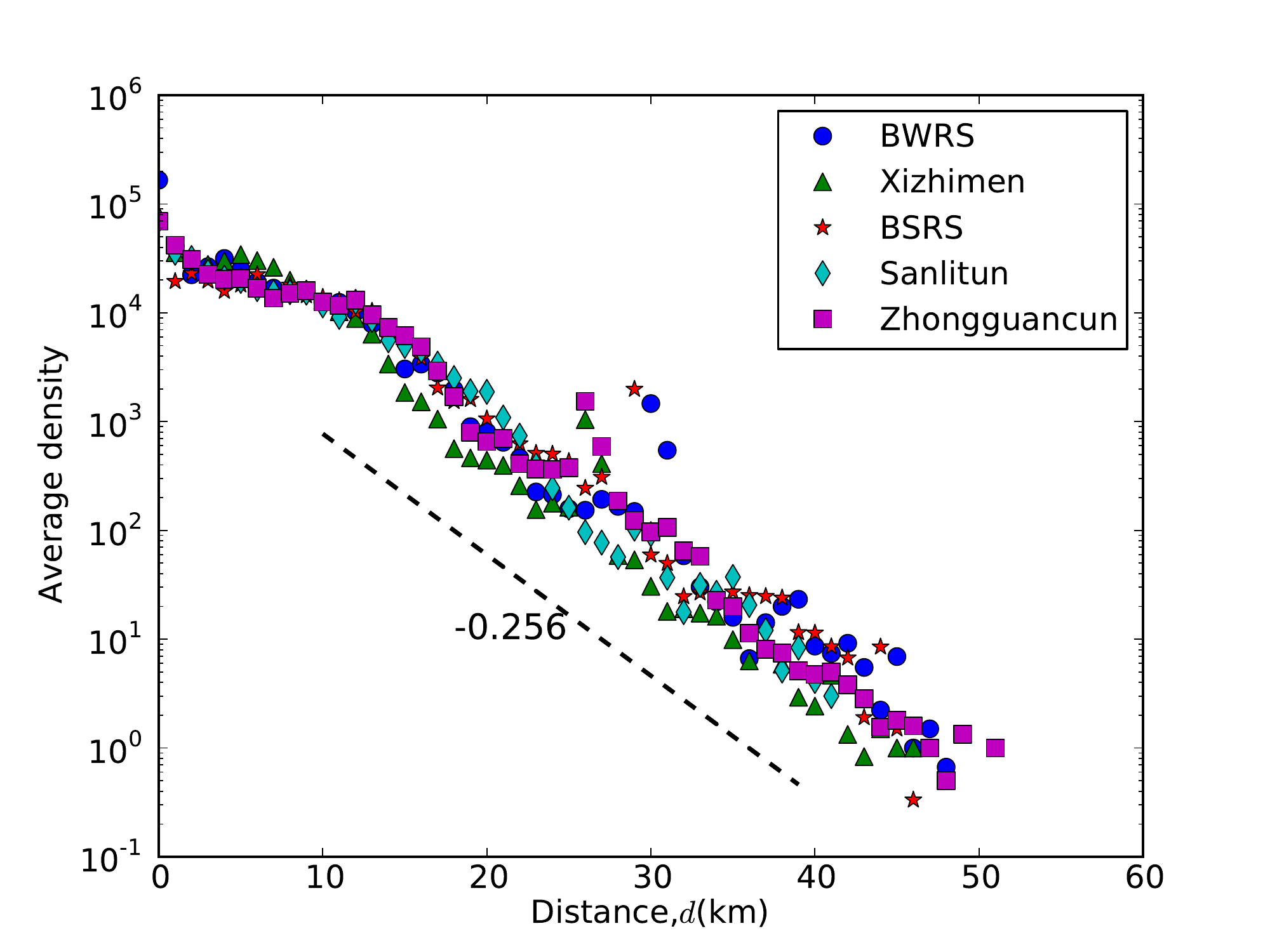}
\label{fig:taxi_dst_decay}
}
\subfloat[Sina Weibo] {
\includegraphics[scale=.4]{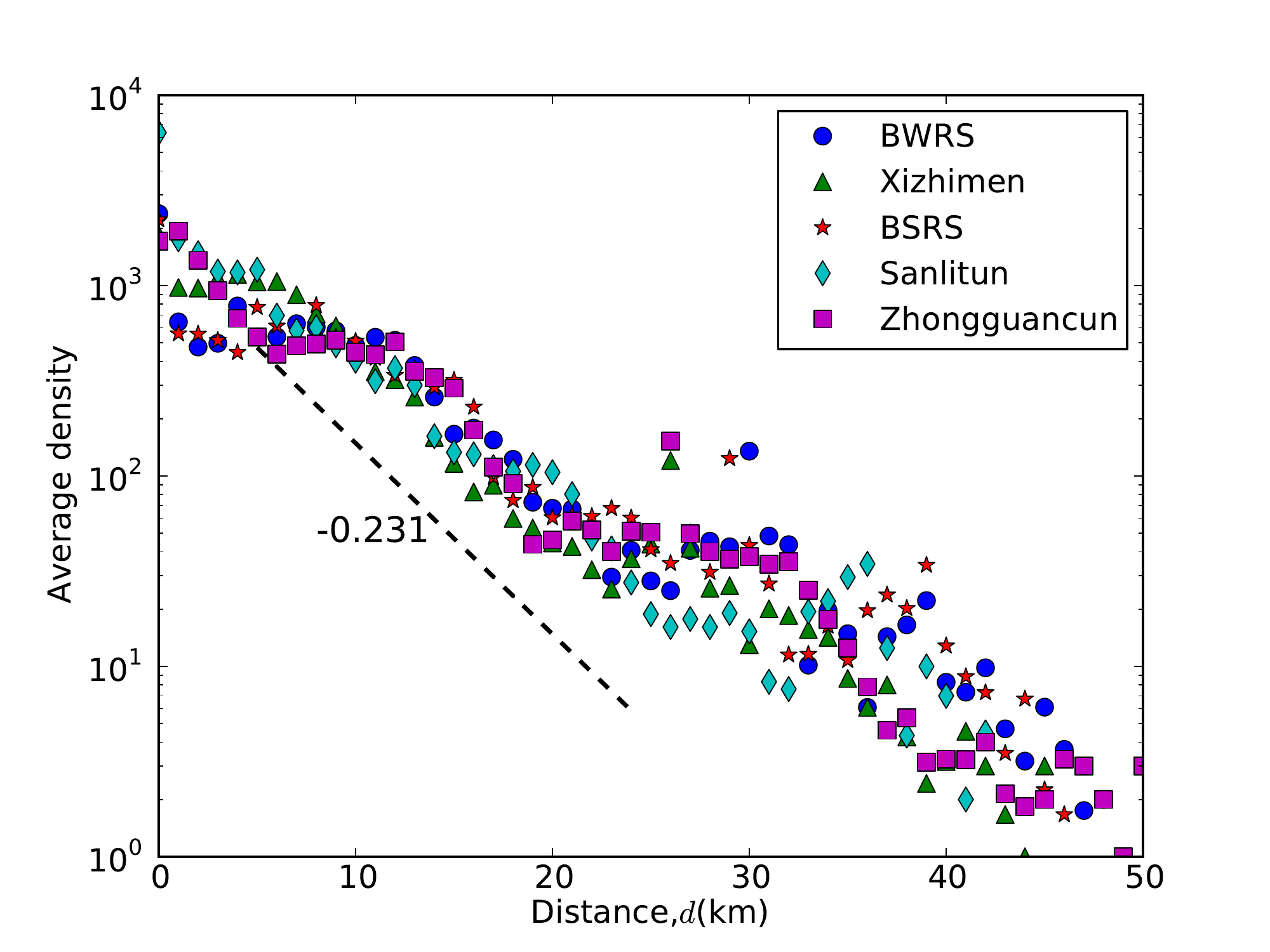}
\label{fig:weibo_dst_decay}
}
\caption{The average densities of destinations and geo-tagged
  microblogs with distance from five hot spots
  regions in taxis and Sina Weibo datasets.}
\label{fig:density}
\end{figure}

As illustrated above, it is no coincidence that exponential law
is discovered in urban areas of cities. Then it is aimed to explore the
origin of exponential law emerged in urban areas. 
Considering five hot spots regions in urban areas: Beijing West
Railway Station (BWRS), Xizhimen, Beijing South Railway Station
(BSRS), Sanlitun and Zhongguancun, the average densities of
destinations or geo-tagged posts' locations with distance from these regions are
plotted in Fig. \ref{fig:density}.  From the
Fig. \ref{fig:taxi_dst_decay}, the average densities for the five hot
spots have similar trends and decay exponentially, in which the
exponent of exponential is -0.256 and is not far from the value -0.230 observed from
distance distribution shown in Fig. \ref{fig:cell_disp}. Also in Fig. \ref{fig:weibo_dst_decay},  the
densities can be fitted by an exponential with exponent -0.231 when
distances lie between 10km and 20km and then decay more
quickly. It is worth mentioning that these findings seem like Clark's \cite{Clark1951}
who first use the negative exponential function to describe urban
population density. These
illustrate that the distributions of destinations and geo-tagged
posts' locations
are very similar near centers of the city and these distributions may
account for the exponential law in urban areas of cities.

\begin{figure}[htbp]
\centering
\includegraphics[scale=.4]{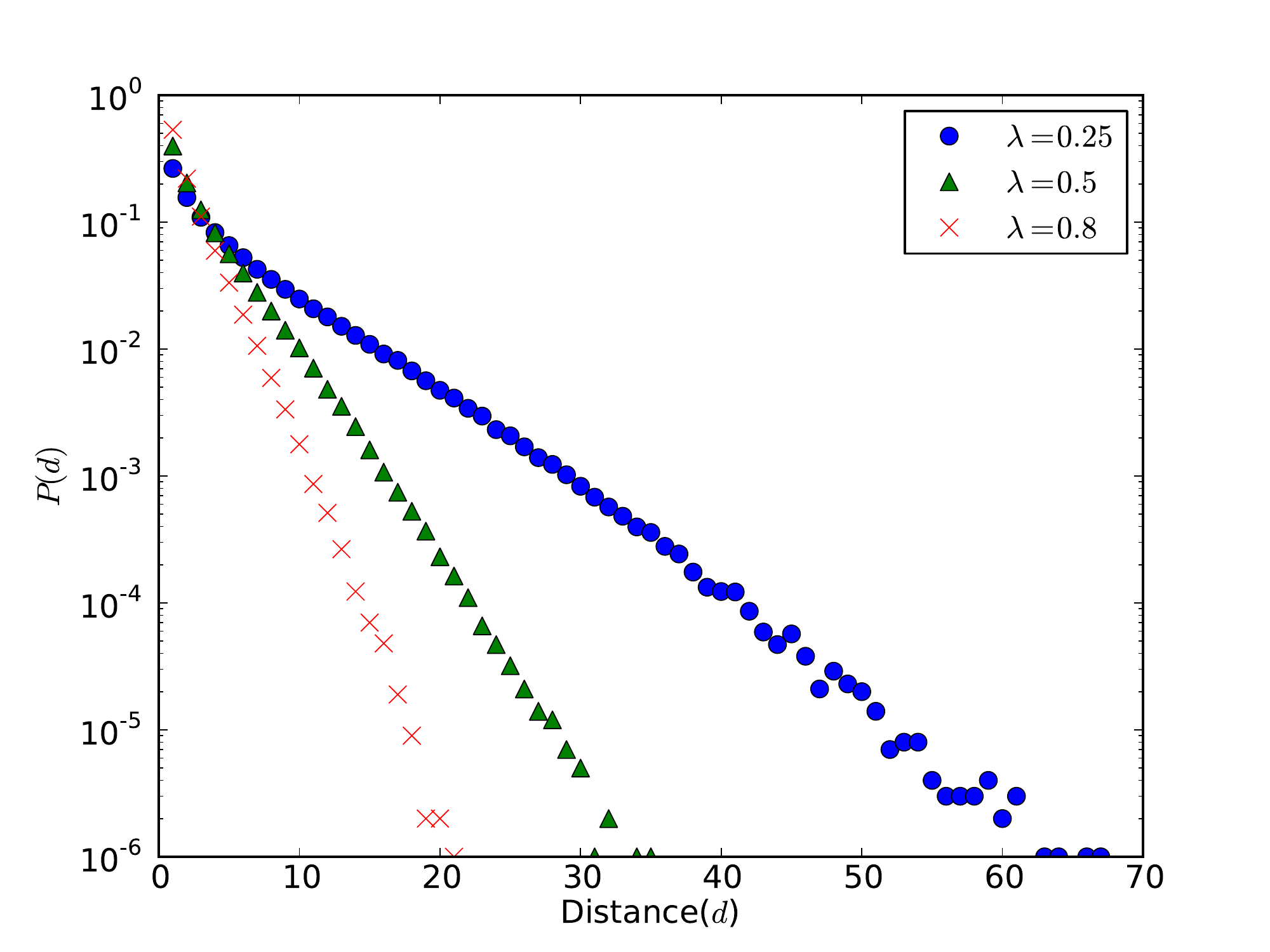}
\caption{Simulations based on our model with average density
    distributions decreasing exponentially for different exponents.}
\label{fig:monocenter_exp}
\end{figure}

Therefore, assuming the density distributions
are described by negative exponential functions for different exponents
$\lambda$, we simulate human travels based on our model with the
parameter $\sigma = 1.601$. 
As shown in Fig. \ref{fig:monocenter_exp}, different exponential exponents
$\lambda$ about 0.25, 0.5 and 0.8 are considered. It is noticed that these
distributions have obvious exponential decreasing trends, where the
fitted exponents of tails of distributions are 0.200, 0.455 and 0.808
separately. As proved in appendix \ref{add:exp}, when the density function is
exponential, the trip distance distribution $P(d)$
satisfies $C_1d^{1-\sigma}e^{-\lambda d} \leq P(d) \leq
C_2d^{1-\sigma}e^{-\lambda d}$. So when $d > 1/\lambda$, the
exponential section dominates and $P(d)$ decreases exponentially. The
fitted exponents are very close to the parameter $\lambda$ of density
function, which accords with our theoretical proof very well.
In urban areas, the density function usually
decreases significantly leading to a large exponent $\lambda$. It
must be noticed that the exponent of $d$ is usually larger than -1 because of the
small parameter $\sigma$ for urban areas, which is different from the observed
power-law distributions of distance where the exponents are between -2
and -1. So it can explain the reason why human trip
distance in urban areas accords with exponential distribution more
better. 

\section{Conclusions and future work}
In our daily life, most human activities, especially movements, are
concentrated in urban areas of cities. So it is very important to
understand intra-city mobility patterns. In the paper, we aim to study
human mobility patterns in urban areas through taxis dataset in Beijing. 

Firstly,  the geographic distributions of origins and destinations
follow very similar patterns. And compared with geo-tagged microblogs'
locations,
they are also very close each other near centers of the city. It suggests that
these distributions are irrelevant to means of transport of human
travels and only depend on urban planning, population distribution and
other properties of the city.

Secondly, it seems that radiation model can not model collective human
movement in urban areas very well. So we propose our model and
observe the exponential law in the distribution of simulated trip
lengths. Furthermore, our model may be appropriate for human travel
not only in urban areas but in countries or larger ranges.

Finally, based on our model, it can be found that the distribution of trip
distances depends on geographic distribution of human travel demands, which is inherent
nature of the city. Meanwhile, it is observed that average human
movement intensities decay exponentially with distance from
hotspots. It can explain the origin of exponential law discovered in
actual trip length distribution.

However, it must be emphasized that intra-urban mobility considered
here occurred during a period of long time. In fact, the traffic flows
between regions in urban areas is varied with the time of a day, which
show strong periodic fluctuations. So in future, we will focus on the
temporal characteristics of intra-urban individual flows to predict
human mobility more precisely.

%% == end of paper:

%% Optional Materials and Methods Section
%% The Materials and Methods section header will be added automatically.

%% Enter any subheads and the Materials and Methods text below.
%\begin{materials}
% Materials text
%\end{materials}

%% Optional Appendix or Appendices
%% \appendix Appendix text...
%% or, for appendix with title, use square brackets:
%% \appendix[Appendix Title]
\appendix
\section{Geo-tagged dataset of Sina Weibo}\label{app:weibo}
Microblogging services, such as Twitter and Weibo, have become more and
more popular for users to share information with friends or
followers. Recently, Weibo, Twitter and other online location-based
services allow users to post their current geographic locations in messages. In
Sina Weibo, when a user posts a geo-tagged microblog, it appears
in a "public timeline" of recent location updates. So by using Weibo's
geolocation API, we monitor the public timeline from Oct. 8, 2012 to
Nov. 4, 2012 and only focus on the microblogs located in urban areas of Beijing.
A total of 513315 geo-tagged posts are collected at first. After
removing abnormal users and repeated microblogs posted by the same user
in a relative short time, we finally obtain 491513 geo-tagged posts.

\section{MLE for our model}\label{add:mle}
Let us consider intra-urban trips during a period of time, which can be denoted
as $Tr = \{(l_O^{(r)}, l_D^{(r)}, t_O^{(r)}, t_D^{(r)}) |
r=1,\ldots,n\}$, where $n$ is the number of trips. Supposing
these trips are independent of each other, 
so the log probability of $Tr$ can be calculated as follows
\begin{eqnarray*}
\log{P(Tr)} & = & \log{\prod_{r=1}^{n}{P(l_O^{(r)}\to l_D^{(r)})}} \\
& = & \sum_{i,j,i\neq j}{T_{ij}\log{P(i\to j)}} \\
& = & \sum_{i,j,i\neq
  j}{T_{ij}\log{\frac{P_O(i)P_D(j)/f(d_{ij})}{\sum_{k\neq
        i}{P_D(k)/f(d_{ik})}}}} \\
& = & \sum_{i,j,i\ne j}{T_{ij} \log{\frac{P_O(i)P_D(j)}{M(i) f(d_{ij})}}}.
\end{eqnarray*}
Here, the Nelder-Mead simplex algorithm \cite{Nelder01011965} is used  to find the
minimum of $-\log{P(Tr)}$ and evaluate the parameter $\sigma$ or
$\lambda$ in the function $f(d)$ numerically.

\section{Proof of uniform density distribution}\label{add:uniform}
Supporting the probability $P(\cdot)$ is uniform
(i.e. $P(i)=1/N$, $(i=1,\ldots,N)$), the probability with travel distance $d$
simulated 
based on our model can be denoted as
\begin{eqnarray*}
P(d) &=& \sum_{i,j:d_{ij} = d}{P(i\to j)}\\
&=& \sum_{i}{P(i)\sum_{j:d_{ij} = d}{\frac{P(j)/d^{\sigma}}{\sum_{k\ne
i}{P(k) / d_{ik}^{\sigma}}}}},
\end{eqnarray*}
where $d_{ij}$ represents the distance between cell $i$ and $j$, and $P(i)$
stands for the probability to select the cell $i$.

When $N$ is large, for different $i$, $\sum_{k\ne i}{P(k) /
d_{ik}^{\sigma}}$ have approximately the same values. Therefore,
\begin{equation*} 
P(d)\propto N\cdot \frac{1}{N} \cdot d / d^{\sigma} \propto d^{1-\sigma}
\end{equation*}

\section{Proof of exponential density distribution}\label{add:exp}

\begin{figure*}[htbp]
\centering
\includegraphics[trim=120 300 10 300]{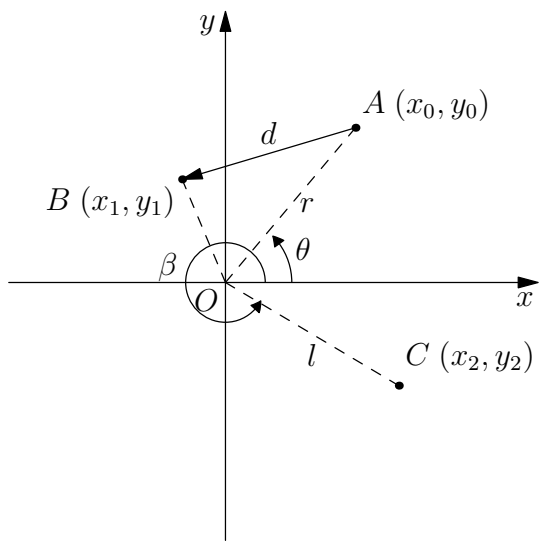}
\caption{\label{fig:monocenter}Illustration of negative exponential density function.}
\end{figure*}

As shown in Fig. \ref{fig:monocenter}, the center is the point $O$ and
the density distribution is 
$$\rho(r) = Ce^{-\lambda r}\quad (0\leq r\leq R),$$
where $r$ is the distance from the center $O$ and $R$ is the size of a
city. By using our model, we can estimate the displacement
distribution of human movement as follows:

$$P(d) = \sum_{d(A,B)=d}{P(A\rightarrow B)}$$

Considering continuity of density distribution, it can also be written
by

$$P(d) = \int\!\!\!\int \rho(A)
\frac{\int_{s:d(A,B)=d}{\rho(B)d^{-\sigma}\mathrm{d}s}}{\int\!\!\!\int
_{C\neq B} \rho(C)d(C,A)^{-\sigma} \mathrm{d}S_2}
\mathrm{d}S_1$$

Let $$I_1 = \int_{s:d(A,B)=d}{\rho(B)d^{-\sigma}\mathrm{d}s}$$
and $$I_2 = \int\!\!\!\int
_{C\neq B} \rho(C)d(C,A)^{-\sigma} \mathrm{d}S_2.$$

Therefore, 
\begin{eqnarray*}
I_1 &=& \int_{s:(x_1-x_0)^2 +
(y_1-y_0)^2=d}{Ce^{-\lambda\sqrt{x_1^2+y_1^2}}d^{-\sigma}\mathrm{d}s}\\ 
& & (x_1=x_0+d\cos{\alpha},y_1=y_0+d\sin{\alpha})\\
&=& \int_0^{2\pi} Cd^{1-\sigma}e^{-\lambda \sqrt{x_0^2 + y_0^2 + d^2 +
2d(x_0\cos{\alpha} + y_0\sin{\alpha})}}\mathrm{d}\alpha,\\
I_2 &=& \int\!\!\!\int
_{(x_2,y_2)\neq(x_0,y_0)}
\frac{Ce^{-\lambda\sqrt{x_2^2+y_2^2}}}{(\sqrt{(x_2-x_0)^2+(y_2-y_0)^2})^\sigma}
\mathrm{d}x_2\mathrm{d}y_2\\
& & (x_2=l\cos{\beta}, y_2=l\sin{\beta})\\
&=& \int_0^{2\pi} \mathrm{d}\beta \int_{0}^{R} \frac{Cle^{-\lambda
l}}{(\sqrt{l^2+x_0^2+y_0^2-2l(x_0\cos{\beta}+y_0\sin{\beta})})^{\sigma}}\mathrm{d}l.\\ 
\end{eqnarray*}

$P(d)$ can be represented as
\begin{eqnarray*}
P(d) &=& \int\!\!\!\int
Ce^{-\lambda\sqrt{x_0^2+y_0^2}}\frac{I_1}{I_2}\mathrm{d}x_0\mathrm{d}y_0\\
& & (x_0=r\cos{\theta},y_0=r\sin{\theta})\\
&=& \int_0^{2\pi} \mathrm{d}\theta \int_0^R Cre^{-\lambda r}\frac{U}{V}\mathrm{d}r
%\frac{\int_0^{2\pi} d^{1-\sigma}e^{-\lambda \sqrt{r^2+d^2+2dr\cos{(\theta-\alpha)}}}\mathrm{d}\alpha}{\int_0^{2\pi} \mathr%m{d}\beta \int_{0}^{R} le^{-\lambda
%l}{(\sqrt{l^2+r^2-2rl\cos{(\beta-\theta)}})^{-\sigma}}\mathrm{d}l}\mathrm{d}r
\end{eqnarray*}
Where
\begin{eqnarray*}
U &=& \int_0^{2\pi} d^{1-\sigma}e^{-\lambda
  \sqrt{r^2+d^2+2dr\cos{(\theta-\alpha)}}}\mathrm{d}\alpha,\\
V &=& \int_0^{2\pi} \mathrm{d}\beta \int_{0}^{R} le^{-\lambda
l}{(\sqrt{l^2+r^2-2rl\cos{(\beta-\theta)}})^{-\sigma}}\mathrm{d}l.
\end{eqnarray*}

We notice that the denominator $V$ has nothing to do with $d$, so consider
the numerator $U$
\begin{eqnarray*}
U &=& \int_0^{2\pi} d^{1-\sigma}e^{-\lambda
\sqrt{r^2+d^2+2dr\cos{(\theta-\alpha)}}}\mathrm{d}\alpha\\
&\geq & \int_0^{2\pi} d^{1-\sigma}e^{-\lambda (r+d)}\mathrm{d}\alpha\\
&=& 2\pi d^{1-\sigma} e^{-\lambda d} e^{-\lambda r}.
\end{eqnarray*}

In a similar way, 
\begin{eqnarray*}
U &\leq& \int_0^{2\pi} d^{1-\sigma}e^{-\lambda \lvert r-d
\rvert}\mathrm{d}\alpha\\
&\leq& \int_0^{2\pi} d^{1-\sigma}e^{-\lambda (d-r)}\mathrm{d}\alpha\\
&=& 2\pi d^{1-\sigma} e^{-\lambda d} e^{\lambda r}.
\end{eqnarray*}

As a result, 

$$C_1d^{1-\sigma}e^{-\lambda d} \leq P(d) \leq
C_2d^{1-\sigma}e^{-\lambda d}$$
\label{add:monocenter}

%% PNAS does not support submission of supporting .tex files such as BibTeX.
%% Instead all references must be included in the article .tex document. 
%% If you currently use BibTeX, your bibliography is formed because the 
%% command \verb+\bibliography{}+ brings the <filename>.bbl file into your
%% .tex document. To conform to PNAS requirements, copy the reference listings
%% from your .bbl file and add them to the article .tex file, using the
%% bibliography environment described above.  

%%  Contact pnas@nas.edu if you need assistance with your
%%  bibliography.

% Sample bibliography item in PNAS format:
%% \bibitem{in-text reference} comma-separated author names up to 5,
%% for more than 5 authors use first author last name et al. (year published)
%% article title  {\it Journal Name} volume #: start page-end page.
%% ie,
% \bibitem{Neuhaus} Neuhaus J-M, Sitcher L, Meins F, Jr, Boller T (1991) 
% A short C-terminal sequence is necessary and sufficient for the
% targeting of chitinases to the plant vacuole. 
% {\it Proc Natl Acad Sci USA} 88:10362-10366.

%% Enter the largest bibliography number in the facing curly brackets
%% following \begin{thebibliography}

%\bibliographystyle{plain}
%\bibliography{paper-reference}

\end{document}